\documentclass{article}%
\usepackage{amsfonts}
\usepackage{amsmath}
\usepackage{amssymb}
\usepackage{graphicx}%
\providecommand{\U}[1]{\protect\rule{.1in}{.1in}}
\newtheorem{theorem}{Theorem}

\newtheorem{example}{Example}

\newtheorem{lemma}{Lemma}

\newtheorem{remark}{Remark}

\begin{document}

\title{M-estimators for Isotonic Regression}
\author{Enrique E. \'{A}lvarez$^{1}$ and V\'{\i}ctor J. Yohai$^{2}$\\$^{1}$University of La Plata and CONICET\\$^{2}$University of Buenos Aires and CONICET}
\date{~}
\maketitle

\begin{abstract}
In this paper we propose a family of \ robust estimates for isotonic
regression: isotonic M-estimators. We show that their asymptotic distribution
is, up to an scalar factor, the \ same as that of Brunk 's \ classical
isotonic estimator. We also derive the influence function and the breakdown
point of these estimates. Finally we perform a Monte Carlo study that shows
that the proposed family includes estimators that are simultaneously highly
efficient under gaussian errors and highly robust when the error distribution
has heavy tails. \ 

\medskip

\end{abstract}

\section{Introduction}

Let $x_{1},\ldots,x_{n}$ be independent random variables collected along
observation points $t_{1}\leq\ldots\leq t_{n}$ according to the model
\begin{equation}
x_{j}=\mu(t_{j})+u_{j}, \label{eqn:model}%
\end{equation}
where the $u_{j}$'s are i.i.d.\ symmetric random variables with distribution
$G$. In \textit{isotonic regression} the trend term $\mu(t)$ is monotone
non-decreasing, i.e., $\mu(t_{1})\leq\ldots\leq\mu(t_{n})$, but it is
otherwise arbitrary. In this set-up, the classical estimator of $\mu(t)$ is
the function $g$ which minimizes the $L_{2}$ distance between the vector of
observed and fitted responses, i.e, it minimizes,
\begin{equation}
\sum_{j=1}^{n}[x_{j}-g(t_{j})]^{2} \label{eqn:L2}%
\end{equation}
in the class $\mathcal{G}$ of non-decreasing piecewise continuous functions.
It is trivial but noteworthy that Equation (\ref{eqn:L2}) posits a finite
dimensional convex constrained optimization problem. Its solution was first
proposed by Brunk (1958) and has received extensive attention in the
Statistical literature (see e.g., Robertson, Wright and Dyskra (1988) for a
comprehensive account). It is also worth noting that any piecewise continuous
non-decreasing function which agrees with the optimizer of (\ref{eqn:L2}) at
the $t_{j}$'s will be a solution. For that reason, in order to achieve
uniqueness, it is traditional to restrict further the class $\mathcal{G}_{0}$
to the subset of piecewise constant non-decreasing functions. Another valid
choice consists in the interpolation at the knots with non-decreasing cubic
splines or any other piecewise continuous monotone function, e.g., Meyer
(1996). We will call this estimator the \emph{ L}$_{\emph{2}}$\emph{ isotonic
estimator}.\ 

The sensitivity of this estimator to extreme observations (outliers) was noted
by Wang and Huang (2002), who propose minimizing instead using the $L_{1}$
norm, i.e. , minimizing%
\[
\sum_{j=1}^{n}\left\vert x_{j}-g(t_{j})\right\vert .
\]
This estimator will be call here\emph{ L}$_{1}$ \textit{Isotonic estimator}.
Wang and Huang (2002) developed the asymptotic distribution of the trend
estimator at a given observation point $t_{0}$ and obtained the asymptotic
relative efficiency of this estimator compared with the classical L$_{2}%
$estimator. Interestingly, this efficiency turned out to be $2/\pi=0.637$, the
same as in the i.i.d.\ location problem.

In this paper we will propose instead a robust\ \emph{isotonic M-estimator}
aimed at balancing robustness with efficiency. Specifically we shall seek the
minimizer of
\begin{equation}
\sum_{j=1}^{n}\rho\left(  \frac{x_{j}-g(t_{j})}{\widehat{\sigma}_{n}}\right)
\label{eqn:Mrepres}%
\end{equation}
where $\widehat{\sigma}_{n}$ is a an estimator of the error scale previously
obtained and $\rho$\ satisfies the following properties

\begin{description}
\item[A1] (i) $\rho(x)$ is non-decreasing in $|x|$, (ii) $\rho(0)=0$, (iii)
$\rho$ is even, (iv) $\rho(x)$ is strictly increasing for $x>0$ and (v) $\rho$
has two continuous derivatives and $\psi=\rho^{\prime}$ is bounded and
monotone non-decreasing.
\end{description}

Clearly, the $L_{2}$ choice corresponds to taking $\rho(x)=x^{2}$ while the
$L_{1}$ option is akin to opting for $\rho(x)=|x|$. These two estimators do no
require the scale estimator $\widehat{\sigma}_{n}.$

Note that the class of M-estimators satisfying A1 does not include estimators
with a redescending choice for\textbf{ $\psi$}. We believe that the strict
differentiability conditions on $\rho$ \ required in A1 are not strictly
necessary, but they make the proofs for the asymptotic theory simpler.
Moreover, some functions $\rho$ which are not twice differentiable everywhere
\textbf{such} as $|x|$ or the Huber\textbf{s'} functions defined below in
(\ref{hubfam}) can be approximated by functions satisfying A1.

The asymptotic distribution of the L$_{2}$ isotonic estimators at a given
point was found by Brunk (1970) and Wright (1981) and the one of the L$_{1}$
estimator by Wang (2002). They prove that the distribution of these estimators
conveniently normalized converge to the distribution of the slope at zero of
the greatest convex minorant of the two-sided Brownian Motion with parabolic
drift. In this paper, we prove a similar result for isotonic M-estimators. The
focus of this paper is on estimation of the trend term at a single observation
point $t_{0}$. We do not address the issue of distribution of the whole
stochastic process $\{\hat{\mu}_{n}(t),t\in\mathcal{T}\}$. Recent research
along those lines are given by Kulikova and Lopuha\"{a} (2006) and a related
result with smoothing was also obtained simultaneously in Pal and Woodroofe (2006).

This article is structured as follows. In Section \ref{sec:rire} we propose
the robust isotonic M-estimator. In Section \ref{sec:ad} we obtain the
limiting distribution of the isotonic M-estimator when the error scale is
known. In Section \ref{scaleeq} we prove that \ under general conditions the
M-estimators with estimated scale have the same \ asymptotic distribution than
when the scale is known. In Section \ref{sec:IF} we define an influence
function which measures the sensitivity of the isotonic M-estimator to an
infinitesimal amount of pointwise contamination. In Section \ref{sec:BP} we
calculate the breakdown point \ of the isotonic M-estimators. In Section
\ref{simul} we compare by Monte Carlo simulations the finite sample variances
of the estimators for two error distributions: normal and Student with three
degrees of freedom. In Section \ref{sec:GW} we analyze two real dataset using
The L$_{2}$ and the isotonic M-estimators. Section \ref{Appendix}\ is an
Appendix containing the proofs.

\section{Isotonic M-Estimators}

\label{sec:rire}

In similarity with the classical setup, we consider isotonic M-estimators that
minimize the objective function (\ref{eqn:Mrepres}) within the class
$\mathcal{G}_{0}$ of piecewise constant non-decreasing functions. As in the
L$_{2}$ and L$_{1}$ cases, the isotonic M-estimator is a step function with
knots at (some of) the $t_{j}$'s. In Robertson and Waltman (1968) it is shown
that maximum-likelihood-type estimation under isotonic restrictions can be
calculated via \textit{min-max} formulae. Assume first that we know that the
scale parameter (e.g. , \ the MAD, of the $u_{t}$s) is $\sigma_{0}$. Since we
\ are considering M-estimators with $\psi$ non-decreasing (see A1), they can
be view as the maximum likelihood estimators corresponding to errors with
density
\[
g(u)=\frac{\exp\left(  -\frac{1}{^{\sigma_{0}}}\int_{0}^{u}\psi(v/\sigma
_{0})dv\right)  }{%
%TCIMACRO{\dint \limits_{-\infty}^{\infty}}%
%BeginExpansion
{\displaystyle\int\limits_{-\infty}^{\infty}}
%EndExpansion
\left[  \exp\left(  -\frac{1}{^{\sigma_{0}}}\int_{0}^{u}\psi(v/\sigma
_{0})dv\right)  du\right]  }%
\]
Then we can compute \ the isotonic M-estimator at a point $t$ using the
min-max calculation formulae
\begin{equation}
\hat{\mu}_{n}(t)=\max_{u\leq t}\min_{v\geq t}\hat{\mu}_{n}(u,v)=\min_{v\geq
t}\max_{u\leq t}\hat{\mu}_{n}(u,v), \label{eqn:minmax}%
\end{equation}
where $\hat{\mu}_{n}(u,v)$ is the unrestricted M-estimator which minimizes
\begin{equation}
\sum_{j\in C(u,v)}\rho\left(  \frac{x_{j}-\mu}{\sigma_{0}}\right)  \text{,}
\label{eqn:locmin}%
\end{equation}
where $C(u,v)=\{j:1\leq j\leq n;u\leq t_{j}\leq v\}$. Alternatively, if $\rho$
is convex and differentiable, as we are assuming, the terms $\hat{\mu}%
_{n}(u,v)$ in (\ref{eqn:minmax}) can be represented uniquely as a zero of
\begin{equation}
S_{n}(u,v,\mu)=\sum_{j\in C(u,v)}\psi\left(  \frac{x_{j}-\mu}{\sigma_{0}%
}\right)  . \label{eqn:partialsums}%
\end{equation}
In particular, when $\rho(u)=-\log(g(u))+\log(g(0)),$ where $g$ is a
probability density, the isotonic M-estimator coincides with the maximum
likelihood estimator when is $u$ is assumed to have density $g$. In particular
if $g$ is the N$(0,\sigma_{0}^{2})$ density, the MLE is the M-estimator which
defined by $\rho(u)=u^{2}$ and therefore it coincides with the classical
$L_{2}$ estimator. When $g$ is the density of a double exponential
distribution, the MLE is the M-estimator defined \ by $\rho(u)=|u|,$ and
therefore it coincides with the L$_{1}$ isotonic estimator. In these two cases
the estimators are independent of the value of $\sigma_{0}.$ One popular
family of $\psi$ functions to define M-estimators is the Huber family%
\begin{equation}
\psi_{k}^{H}(u)=\text{sign}(u)\min(|u|,k). \label{hubfam}%
\end{equation}

Clearly, when $\sigma_{0}$ is replaced by $\widehat{\sigma}_{n},$ equations
(\ref{eqn:minmax})-(\ref{eqn:partialsums}) still holds with $\sigma_{0}$
replaced by $\widehat{\sigma}_{n}.$ Since $\psi$ is non-decreasing, the
function $S_{n}(u,v,\mu)$ defined in equation (\ref{eqn:partialsums}) is
non-increasing as a function of $\mu$. This entails the fundamental identities
given below
\begin{align}
S_{n}(u,v,a)>0  &  \text{ if and only if }\hat{\mu}_{n}%
(u,v)>a,\label{eqn:equivg}\\
S_{n}(u,v,a)<0  &  \text{ if and only if }\hat{\mu}_{n}(u,v)<a.
\label{eqn:equivs}%
\end{align}
\ These identities will be very useful in the development of the asymptotic distribution.

\section{Asymptotic Distribution \label{sec:ad}}

In this section we derive the asymptotic distribution of the isotonic
M-estimator $\hat{\mu}_{n}(t_{0})$ of $\mu(t_{0}).$ We first make the sample
size $n$ explicit in the formulation of the model by postulating
\begin{equation}
x_{n,i}=\mu(t_{n,i})+u_{n,i}, \label{eqn:tarray}%
\end{equation}
where the errors $\{u_{n,i},1\leq i\leq n\}$ form a triangular array of
i.i.d.\ random variables with distribution $G$ and $\{t_{n,i},1\leq i\leq n\}$
is a triangular array of observation points. Their exact location is described
by the function $H_{n}(t)=n^{-1}\sum_{i=1}^{n}1(t_{n,i}\leq t)$. The values
$t_{n,j}$ may be fixed or random but we will assume that there exists a
continuous distribution function $H$ \ which has as support a finite closed
interval such that%
\begin{equation}
\sup_{t}|H_{n}(t)-H(t)|=o_{P}(n^{-1/3}). \label{condH}%
\end{equation}
\ Without loss of generality we shall assume in the sequel it is the interval
$[0,1]$. \ 

We will study the asymptotic distribution of $\hat{\mu}_{n}(t_{0})$ where
$t_{0}$ is an interior point of $[0,1].$ The classical L$_{2}$ isotonic
estimator $\hat{\mu}_{n}(t_{0}),$ with $t_{0}$ at the boundary of the support
of $H,$ is known to suffer from the so-called \textit{spiking problem} (e.g.,
Sun and Woodroofe, 1999), i.e., $\hat{\mu}_{n}(t_{0})$ is not even
consistent.\ We further make the following assumptions.

\begin{description}
\item[A2] The function $H$ is continuously differentiable in a neighborhood of
$t_{0}$ with $h(t_{0})=H^{\prime}(t_{0})>0$.

\item[A3] For a fixed $t_{0}$, we assume the function $\mu(t)$ has two
continuous derivatives in a neighborhood of $t_{0}$, and $\mu^{\prime}%
(t_{0})>0$.

\item[A4] The error distribution $G$ has a density $g$ symmetric and
continuous with $g(0)>0$.
\end{description}

We consider first the case where $\sigma_{0}$ is known. Our first aim is to
show that isotonic M-estimation is asymptotically a local problem.
Specifically, we will see in Lemma \ref{lemma:rao1}\ that $\hat{\mu}_{n}%
(t_{0})$ depends only on those $x_{j}$ corresponding to observation points
$t_{j}$ lying in a neighborhood of order $n^{1/3}$ about $t_{0}$. This result
is similar to Prakasa Rao (1969), Lemma 4.1, who stated it in the context of
density estimation. Our treatment here will parallel that of Wright (1981),
who worked on the asymptotics of \ the $L_{2}$ isotonic regression estimator
when the smoothness of the underlying trend function $\mu(\cdot)$ is specified
via the number of its continuous derivatives.

Specifically, since $H^{\prime}(t_{0})>0$ we may choose for an arbitrary $c$
and $n$ sufficiently large, positive numbers $\alpha_{l}(n)$ and $\alpha
_{u}(n)$ for which
\[
H(t_{0})-H(t_{0}-\alpha_{l}(n))=H(t_{0}+\alpha_{u}(n))-H(t_{0})=2cn^{-1/3}.
\]
With this, define the \textit{localized version} of the isotonic M-estimator
as
\begin{equation}
\mu_{n}^{\ast}(t_{0})=\max_{t_{0}-\alpha_{l}(n)<u\leq t_{0}}\ \min_{t_{0}\leq
v<t_{0}+\alpha_{u}(n)}\hat{\mu}_{n}(u,v). \label{eqn:modest}%
\end{equation}
Then we have the following Lemma

\begin{lemma}
\label{lemma:rao1} Assume A1-A4 and (\ref{condH}). Then if $\hat{\mu}%
_{n}(t_{0})$ is defined by (\ref{eqn:minmax}), we have,
\begin{equation}
\lim_{c\rightarrow\infty}\limsup_{n\rightarrow\infty}\mathrm{P}[\hat{\mu}%
_{n}(t_{0})\neq\mu_{n}^{\ast}(t_{0})]=0. \label{eqn:rao}%
\end{equation}

\end{lemma}

Is is also noteworthy that the estimator in Equation (\ref{eqn:modest}) is not
computable, for $\alpha_{l}$ and $\alpha_{u}$ depend on the distribution $H$
which is generally unknown. For computational purposes this implies that the
calculation of these estimators will indeed be global for fixed sample sized.
Lemma \ref{lemma:rao1} is, however, crucial to study the asymptotic properties
of $\hat{\mu}_{n}(t)$.

Given an stochastic process $\{Z(v),-\infty<v<\infty\}$, we denote by
``$\text{slogcm}[Z(t)]$'' the random variable that corresponds to the slope at
zero of the greatest convex minorant of $Z(t).$ The following theorem gives
the asymptotic distribution of $\hat{\mu}_{n}(t_{0})$.

\begin{theorem}
Assume A1-A4 and (\ref{condH}), Let $\hat{\mu}_{n}(t_{0})$ be given by
(\ref{eqn:minmax}), then
\begin{equation}
\left[  \frac{1}{2}\mu^{\prime}(t_{0})H^{\prime}(t_{0})\sigma_{0}^{2}%
\frac{\text{\textrm{\textrm{E}$_{G}$(}}\psi^{2}(u/\sigma))}%
{[\text{\textrm{\textrm{E}$_{G}$}}(\psi^{\prime}(u/\sigma))]^{2}}\right]
^{-1/3}n^{1/3}\left(  \hat{\mu}_{n}(t_{0})-\mu(t_{0})\right)  \Rightarrow
\text{\textrm{slogcm}}\left(  \mathbb{W}(v)+v^{2}\right)  , \label{eqn:slogcm}%
\end{equation}
where $\mathbb{W}(v)$ is a two-sided standard Brownian motion.
\end{theorem}

\begin{remark}
Notice that in the case of the L$_{2}$ isotonic estimator the function
$\rho(x)=x^{2}$, so $\psi(x)=2x$ and $\psi^{\prime}(x)=2$ so that
\textrm{E}$_{G}$($\psi^{\prime}(u))=2$ and $\sigma_{\psi}^{2}=\sigma_{2u}%
^{2}=4\sigma^{2}$. Then the standardizing constant is given by
\[
\frac{1}{2}\mu^{\prime}(t_{0})H^{\prime}(t_{0})\frac{\text{\textrm{E}$_{G}$%
}(\psi^{2}(u))}{(\text{\textrm{E}$_{G}$}(\psi^{\prime}(u)))^{2}}=\frac{1}%
{2}\mu^{\prime}(t_{0})H^{\prime}(t_{0})\sigma^{2},
\]
as it is known for the L$_{2}$ isotonic estimator.
\end{remark}

\begin{remark}
In the case of L$_{1}$ isotonic regression notice that in the function
$\rho(x)=|x|$, so $\psi(x)=$sign$(x)$ for $x\neq0$ or else is left undefined.
Our method is thus not applicable as the assumptions on $\psi$ do not hold.
However, consider a sequence of functions $\psi_{m}(x)$ for which
\begin{equation}
\psi_{m}(x)=%
\begin{cases}
-1 & x\leq-1/m\\
mx & -1/m+1/m^{2}<x<1/m-1/m^{2}\\
1 & x\geq1/m
\end{cases}
, \label{eqn:approxpsi}%
\end{equation}
and so that there is continuity of the first 3 derivatives everywhere; for a
construction of such type of functions it is enough to consider quartic
splines (e.g., De Boor, 2001). In this setup we get
\begin{align*}
\lim_{m\rightarrow\infty}\text{E}_{G}(\psi_{m}^{2}(u))  &  =1\\
\lim_{m\rightarrow\infty}\text{E}_{G}\psi_{m}^{\prime}(u)  &  =\lim
_{m\rightarrow\infty}m\left[  G(1/m-1/m^{2})-G(-1/m+1/m^{2})\right]
=2G^{\prime}(0).
\end{align*}
Letting $m\rightarrow\infty$ and $n\rightarrow\infty$ so that $m/n\rightarrow
\infty$, we obtain
\[
\lim_{n\rightarrow\infty}\frac{1}{2}\mu^{\prime}(t_{0})H^{\prime}(t_{0}%
)\frac{\text{E}_{G}(\psi_{n}^{2}(u))}{(\text{E}_{G}(\psi_{n}^{\prime}%
(u)))^{2}}=\frac{1}{8}\frac{\mu^{\prime}(t_{0})}{[G^{\prime}(0)]^{2}}%
H^{\prime}(t_{0}),
\]
as it is known in the case of L$_{1}$ isotonic regression (see Wang and Huang, 2002).
\end{remark}

\begin{remark}
A similar construction to Equation (\ref{eqn:approxpsi}) may be applied to the
functions $\psi_{k}^{H}$ in the Huber's family.
\end{remark}

\section{Robust Isotonic M-Estimators with a Previous Scale Estimator
\label{scaleeq}}

We will consider now the more realistic case where $\sigma_{0}$ is not known
and it is replaced by an estimator $\widehat{\sigma}_{n}$ previously
calculated. Then, in order to obtain an scale equivariant estimator we should
replace $\sigma_{0}$ in (\ref{eqn:locmin}) and (\ref{eqn:partialsums}) by a
robust scale equivariant estimator $\widehat{\sigma}_{n}.$ In Remarks
\ref{rescalem} and \ref{rescalemed} below we give some possible choices for
$\widehat{\sigma}_{n}.$

In the next Theorem it is shown that under suitable regularity conditions, it
can be proved that if $\widehat{\sigma}_{n}$ converges to $\sigma_{0}$ fast
enough, both isotonic M-estimators, the one using the fixed scale $\sigma_{0}$
and the one using the scale $\widehat{\sigma}_{n}$, have the same asymptotic
distribution.\ Making explicit the scale in the notation, denote the isotonic
M-estimator of $\mu(t)$ based on a fixed scale $\sigma$ by $\hat{\mu}%
_{n}(t,\sigma).$ Then%
\[
\hat{\mu}_{n}(t,\sigma)=\min_{u\leq t}\max_{v\geq t}\hat{\mu}_{n}%
(u,v,\sigma)=\max_{u\leq t}\min_{v\geq t}\hat{\mu}_{n}(u,v,\sigma),
\]
where $\hat{\mu}_{n}(u,v,\sigma)$ solves
\begin{equation}
S_{n}(u,v,\sigma):=\sum_{j\in C(u,v)}\psi\left(  \dfrac{x_{j}-\hat{\mu}%
_{n}(u,v,\sigma)}{\sigma}\right)  =0 \label{eqn:zsigma}%
\end{equation}
over $C(u,v):=\{j:1\leq j\leq n;u\leq t_{j}\leq v\}$.

We need the following Additional Assumptions:

\begin{description}
\item[A5] There exists $k>0$ such that $\psi^{\prime}(u)>0$ for $|u|<k$ and
$\psi^{\prime}(u)=0$ if $|u|>k.$

\item[A6] The estimator $\hat{\sigma}_{n}$\ satisfies $n^{1/3}(\hat{\sigma
}_{n}-\sigma_{0})=o_{P}(1).$
\end{description}

Then we have the following Theorem:

\begin{theorem}
\label{scaleest} Assume A1-A6 \ Then
\[
n^{1/3}|\hat{\mu}_{n}(t,\sigma_{0})-\hat{\mu}_{n}(t,\hat{\sigma}_{n}%
)|=o_{P}(1).
\]
Assume also that (\ref{condH}) holds, then both estimators have the same
asymptotic distribution.
\end{theorem}

\begin{remark}
\label{rescalem}In the context of nonparametric regression Ghement, Ruiz and
Zamar (2008) propose to use as scale estimator $\widehat{\sigma}_{n}$ given
by
\[
\widehat{\sigma}_{n}=\frac{1}{\sqrt{2}}s(x_{2}-x_{1},...,x_{n}-x_{n-1}),
\]
where $s$ is an M-estimator of scale, i.e., $s(u_{1},...,u_{n})$ is defined as
the value $s$ satisfying
\begin{equation}
\frac{1}{n}\sum_{i=1}\chi\left(  \frac{u_{i}}{s}\right)  =b \label{scaleM}%
\end{equation}
where $\chi(u)$ is a function \ which is even, non-decreasing for $u$ $\geq0,$
bounded and continuous. The right hand side is generally taken so that if $u$
is N(0,1), $E\chi\left(  u\right)  =b.$ This condition makes the estimator
converging to the standard deviation when applied to a random sample of the
N(0,1) distribution. A popular family of functions\ $\chi$ to compute scale
M-estimators is the bisquare family given by
\begin{equation}
\chi_{c}(u)=\left\{
\begin{array}
[c]{ccc}%
1-\left(  1-(\frac{u}{c})^{2}\right)  ^{3} & \text{if} & |u|\leq c,\\
1 & \text{if} & |u|>c.
\end{array}
\right.  \label{BISQro}%
\end{equation}
\ Ghement et al. (2008) prove that if $\mu(t)$ is continuous \ under general
conditions on $\chi$ Condition A6 is satisfied \ with $\sigma_{0}$ defined by%
\begin{equation}
\mathrm{E}_{G}\chi_{c}\left(  \frac{u}{\sigma_{0}}\right)  =b. \label{limSsca}%
\end{equation}
$\ $
\end{remark}

\begin{remark}
\label{rescalemed} An alternative scale estimator, which does not require the
continuity of $\mu,$ is provided by%
\[
\widehat{\sigma}_{n}=\frac{1}{\Phi^{-1}(3/4)}\text{median}(|\widehat{u}%
_{1}|,...,|\widehat{u}_{n}|)
\]
where $\widehat{u}_{1},...,\widehat{u}_{n}$ are the residuals corresponding to
the L$_{1}$isotonic\ estimator. We conjecture but we do not have a proof that
this estimator converges also with rate $n^{-1/2}$ to $\sigma_{0}=$%
median$_{G}(|u|)/\Phi^{-1}(3/4).$
\end{remark}

\section{Influence Function \label{sec:IF}}

In order to obtain the influence function of the isotonic M-estimator at a
given point $t$ we need to assume that the pair $(x,t)$ is random. In this
case the isotonic regression model assumes that $x=\mu(t)+u,$ where $u$ is
independent of $t$ and $\mu(t)$ is non-decreasing.\ We assume that the error
term $u$ has a symmetric density $g$, and that the observation point $t$ has a
distribution with density $h$.

We start assuming that $\sigma_{0}$ is known and suppose that we want to
estimate $\mu(t_{0}).$ Given an arbitrary distribution $\Lambda$ of $(t,x)$,
the isotonic \textit{M-estimating functional} of $\mu(t_{0})$ which we
henceforth denote by $T_{t_{0}}(\Lambda)$ is defined in three steps as
follows. First for $r,s\geq0$ let $m(t_{0},r,s,\Lambda)$ be defined \ as the
value $m$ satisfying
\[
\int\limits_{-\infty}^{\infty}\int_{t_{0}-r}^{t_{0}+s}\psi\left(  \frac
{(x-m)}{\sigma_{0}}\right)  d\Lambda=0.
\]
\ \ Let
\[
m^{-}(t_{0},r,\Lambda)=\min_{s\geq0}m(t_{0},r,s,\Lambda),
\]
\ and then $T_{t_{0}}(\Lambda)$ is defined by
\[
T_{t_{0}}(\Lambda)=\max_{r\geq0}m^{-}(t_{0},r,\Lambda).
\]
Let $\Lambda_{n}$ be the empirical distribution of $\{(x_{n,j},t_{n,j}),1\leq
j\leq n\}$, then if $\hat{\mu}_{n}(t)$ is the estimator defined in
(\ref{eqn:minmax}), we have
\[
\hat{\mu}_{n}(t)=T_{t}(\Lambda_{n}).
\]

It is immediate that if $\Lambda_{0}$ is the joint distribution corresponding
to model (\ref{eqn:model}) we have $T_{t_{0}}(\Lambda_{0})=\mu(t_{0}),$ so
that the isotonic M-estimator is Fisher-consistent. Consider now the
contaminated distribution
\[
\Lambda_{\varepsilon,t^{\ast},x^{\ast}}=(1-\varepsilon)\Lambda_{0}%
+\varepsilon\delta_{(t^{\ast},x^{\ast})},
\]
where $\delta_{(t^{\ast},x^{\ast})}$ represents a point mass at $(t^{\ast
},x^{\ast})$. In this case we define the influence function of $T_{t_{0}}$ by
\begin{equation}
\text{IF}^{\ast}(T_{t_{0}},t^{\ast},x^{\ast})=\lim_{\varepsilon\rightarrow
0}\frac{\left(  T_{t_{0}}(\Lambda_{\varepsilon,t^{\ast}x^{\ast}})-T_{t_{0}%
}(\Lambda_{0})\right)  ^{2}}{\varepsilon}. \label{eqn:cuadinf}%
\end{equation}
\ Then, we have the following Theorem:

\begin{theorem}
\label{InfFun}Consider the isotonic regression model given in (\ref{eqn:model}%
) and let $T_{t_{0}}$ be an isotonic M-estimating functional, \ where $t_{0}$
is an interior observation point. Then, under assumptions A1-A4 we have%
\begin{equation}
\text{IF}^{\ast}(T_{t_{0}},t^{\ast},x^{\ast})=\left\{
\begin{array}
[c]{ccc}%
\dfrac{2\mu^{\prime}(t_{0})\sigma_{0}\left\vert \psi((x-\mu(t_{0}))/\sigma
_{0})\right\vert }{h(t_{0})\text{\textrm{E}$_{G}$}(\psi^{\prime}(u/\sigma
_{0}))} & \text{if} & t^{\ast}=t_{0},\\
0 & \text{if} & t^{\ast}\neq t_{0}.
\end{array}
\right.  \label{IF*}%
\end{equation}

\end{theorem}

Notice that in the numerator of (\ref{eqn:cuadinf}) appears the square of the
bias instead of the plain bias as in the classical definition of Hampel
(1974). Therefore for the isotonic M-estimator $T_{t_{0}}$ the bias caused by
a point mass contamination $(t_{0},x^{\ast})$ is of order $\varepsilon^{1/2}$
instead of the usual order of $\varepsilon$.\ 

Alternatively, it is also of interest to know what happens when we are
estimating $\mu(t_{0})$ and contamination takes place at a point $t^{\ast}\neq
t_{0}.$ According to (\ref{IF*}), the influence function in this case is zero.
This occurs because in this case for $\varepsilon$ sufficiently small
$T_{t_{0}}(\Lambda_{\varepsilon,t^{\ast}x^{\ast}})=T_{t_{0}}(\Lambda_{0})$.

It is easy to show that when we use a scale $\widehat{\sigma}_{n}%
\rightarrow\sigma_{0}$ defined by a continuous functional, the influence
function of the isotonic M-estimator is still given by (\ref{IF*}).

\section{Breakdown Point \label{sec:BP}}

Roughly speaking the breakdown point of an estimating functional $T_{t_{0}}$
of $\mu(t_{0})$ is the smallest fraction of outliers which suffices to drive
$|T_{t_{0}}|$ to infinity. More precisely, consider the contamination
neighborhood $\mathcal{V}_{\Lambda_{0},\varepsilon}$ of the distribution
$\Lambda_{0}$ of size $\varepsilon$ defined as
\[
\mathcal{V}_{\Lambda_{0},\varepsilon}=\{\Lambda:\Lambda=(1-\varepsilon
)\Lambda_{0}+\varepsilon\Lambda^{\ast}\},
\]
where $\Lambda^{\ast}$ is an arbitrary distribution of $(x,t)$ such that $t$
takes values in $[0,1]$ and $x$ in $\mathbb{R}$. The asymptotic breakdown
point of $T_{t_{0}}$ at $\Lambda_{0}$ is defined by
\[
\varepsilon^{\ast}(T_{t_{0}},\Lambda_{0})=\inf\left\{  \varepsilon
:\sup_{\Lambda\in\mathcal{V}_{\Lambda_{0}\varepsilon}}\left\vert T_{t_{0}%
}(\Lambda)\right\vert =\infty\right\}  .
\]
We start considering the case that $\sigma_{0}$ is known. Then we have the
following theorem.

\begin{theorem}
Consider the isotonic regression model given in (\ref{eqn:model}) and let
$T_{t_{0}}$ be an isotonic M-estimating functional where $t_{0}$ is an
interior observation point. Then under assumptions A1-A4 we have%
\[
\varepsilon^{\ast}(T_{t_{0}},\Lambda_{0})\geq\min\left\{  \dfrac{H(t_{0}%
)}{1+H(t_{0})},\dfrac{1-H(t_{0})}{2-H(t_{0})}\right\}  .
\]
In the special case when $H$ is uniform, this becomes
\begin{equation}
\varepsilon^{\ast}(T_{t_{0}},\Lambda_{0})\geq\min\left\{  \dfrac{t_{0}%
}{1+t_{0}},\dfrac{1-t_{0}}{2-t_{0}}\right\}  \label{eqn:breakunif}%
\end{equation}
which takes a maximum value of $1/3$ at $t_{0}=1/2.$\ 
\end{theorem}

In the case that $\sigma_{0}$ is replaced by an estimator $\widehat{\sigma
}_{n}$ derived from a \ continuous functional $S,$ it can be proved that the
breakdown point of $T_{t_{0}}$ satisfies
\[
\varepsilon^{\ast}(T_{t_{0}},\Lambda_{0})\geq\min\left\{  \dfrac{H(t_{0}%
)}{1+H(t_{0})},\dfrac{1-H(t_{0})}{2-H(t_{0})},\varepsilon^{\ast}(\Lambda
_{0})\right\}  .
\]

Ghement et al. (2008) showed that if $\widehat{\sigma}_{n}$ is defined as in
Remark \ref{rescalem}, where $s$ is defined by (\ref{scaleM})-(\ref{limSsca})
with $c=0.7094$ and $b=3/4,$then $\varepsilon^{\ast}(\Lambda_{0})=0.5$ .
Moreover in this case $\sigma_{0}$ coincides with the standard deviation when
the error has a normal distribution.

\section{Examples\label{sec:GW}}

\begin{example}
\label{inmorTA}In this section we consider data on Infant Mortality across
Countries. The dependent variable, the number of infant deaths per each
thousand births is assumed decreasing in the country's per capita income.
These data are part of the R package \textquotedblleft
faraway\textquotedblright\ and was used in Faraway (2004). The manual of this
package only mentions that the data are not recent but it does not give
information on the year and source. In Figure \ref{morinf} we compare the
L$_{2}$ isotonic regression estimator with the isotonic M-estimator computed
with the Huber's function with $k=0.98$ \ and $\widehat{\sigma}_{n}$ as in
Remark 2, where $s$ is defined by (\ref{scaleM})-(\ref{limSsca}) with
$c=0.7094$ and $b=3/4.$ There are four countries with mortality above 250:
Saudi Arabia (650), Afghanistan (400), Libya (300) and Zambia (259). These
countries, specially Saudi Arabia and Libya due to their higher relative
income per capita, exert a large impact on the L$_{2}$ estimator. The robust
choice, on the other hand, appears to resistant to these outliers and provides
a good fit.
\end{example}

%

%TCIMACRO{\FRAME{ftbpFU}{6.6322in}{3.5993in}{0pt}{\Qcb{Infant Mortality Data.
%The solid line corresponds to the classical isotonic regression and \ the
%dashed line to the isotonic M-estimate}}{\Qlb{morinf}}{morewm.emf}%
%width 6.6322in;  height 3.5993in;  depth 0pt;  original-width 14.9111in;
%original-height 8.0635in;  cropleft "0";  croptop "1";  cropright "1";
%cropbottom "0";  filename 'morewm.EMF';file-properties "XNPEU";}}}%
%BeginExpansion
\begin{figure}
[ptb]
\begin{center}
\includegraphics[
natheight=8.063500in,
natwidth=14.911100in,
height=3.5993in,
width=6.6322in
]%
{morewm.emf}%
\caption{Infant Mortality Data. The solid line corresponds to the classical
isotonic regression and \ the dashed line to the isotonic M-estimate}%
\label{morinf}%
\end{center}
\end{figure}
%EndExpansion

\begin{example}
We reconsider the Global Warming dataset first analyzed in the context of
isotonic regression by Wu, Woodroofe and Mentz~(2001) from a classical
perspective and subsequently analyzed from a Bayesian perspective in Alvarez
and Dey (2009). The original data is provided by Jones \textit{et al.} (see
http//cdiac.esd.ornl.gov/trends/temp/jonescru/jones.html) containing annual
temperature anomalies from 1858 to 2009, expressed in degrees Celsius and are
relative to the 1961-1990 mean. Even though the global warming data, being a
time series, might be affected by serial correlation, e.g. Fomby and
Vogelsang~(2002), we opted for simplicity as an illustration to ignore that
aspect of the data and model it as a sequence of i.i.d.\ observations.
%TCIMACRO{\FRAME{ftbpFU}{4.2341in}{4.2424in}{0pt}{\Qcb{World Annual Weather
%Anomalies}}{\Qlb{fig:gw1}}{gw3.eps}{\special{ language "Scientific Word";
%type "GRAPHIC";  maintain-aspect-ratio TRUE;  display "USEDEF";
%valid_file "F";  width 4.2341in;  height 4.2424in;  depth 0pt;
%original-width 8.89in;  original-height 8.9066in;  cropleft "0";
%filename 'gw3.eps';file-properties "XNPEU";}}}%
%BeginExpansion
\begin{figure}
[ptb]
\begin{center}
\includegraphics[
height=4.2424in,
width=4.2341in
]%
{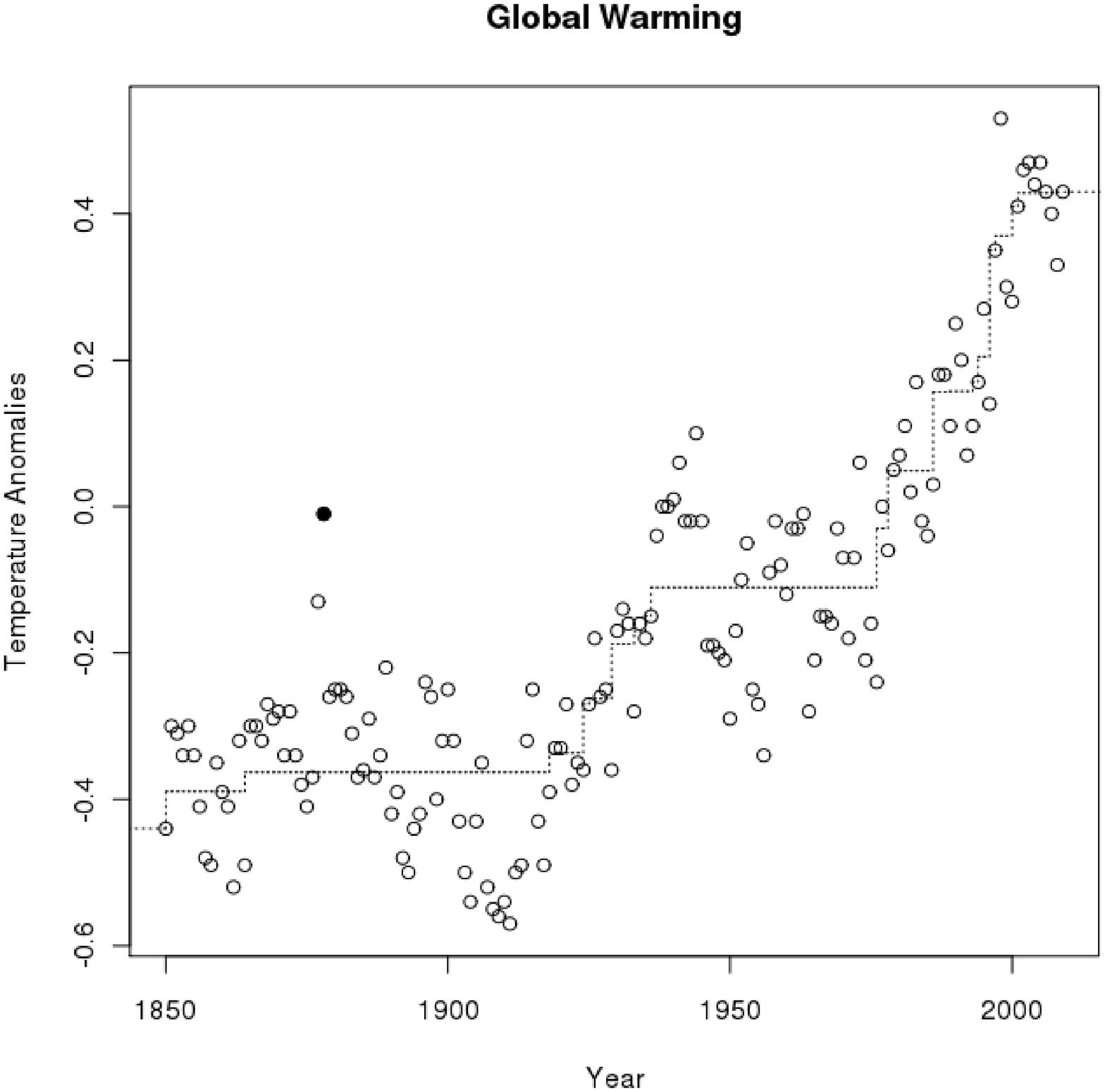}%
\caption{World Annual Weather Anomalies}%
\label{fig:gw1}%
\end{center}
\end{figure}
%EndExpansion

\end{example}

\textit{In Figure \ref{fig:gw1} we plot the L}$_{2}$\textit{ isotonic
estimator, which for these data is identical to the isotonic M-estimate with
k=0.98. Visual inspection of the plot shows a moderate outlier corresponding
to the year 1878 (shown as a solid circle). That apparent outlier, however,
has no effect on the estimator due to the isotonic character of the
regression.The fact that the L}$_{2}$\textit{ and the isotonic M-estimates
coincide for these data seems to indicate that the phenomenon of Global
Warming is not due to isolated outlying anomalies, but it is due instead to a
steady increasing trend phenomenon. In our view, that validates from the point
of view of robustness, the conclusions of other authors on the same data
(e.g.\ Wu, Woodroofe and Mentz (2001), and \'{A}lvarez and Dey (2009)) who
have rejected the hypothesis of constancy in series of the worlds annual
temperatures in favor of an increasing trend.}

\section{Monte Carlo results \label{simul}}

Interestingly the limiting distribution of the Isotonic \textit{M}-estimator
is based on the ratio
\[
\frac{\mathrm{\mathrm{E}_{G}}(\psi^{\prime}(u/\sigma_{0}))^{2}}%
{\mathrm{\mathrm{E}_{G}}(\psi^{2}(u/\sigma_{0}))}%
\]
as in the the i.i.d.\ location problem (e.g.\ Maronna, Martin and Yohai,
2006). The slower convergence rate, however, entails that the respective
asymptotic relative efficiencies are those of the location situation taken to
the power $2/3$. Specifically, note that from Theorem 1 for any isotonic
M-estimator
\begin{align}
&  \text{avar}\left\{  n^{1/3}\left[  \hat{\mu}_{n}(t_{0})-\mu(t_{0})\right]
\right\} \\
&  =\left[  \frac{1}{2}\mu^{\prime}(t_{0})H^{\prime}(t_{0})\frac
{\mathrm{E}_{G}[\psi(u)^{2}]}{[\mathrm{E}_{G}\psi^{\prime}(u)]^{2}}\right]
^{2/3}\text{var}[\text{slogcm}\left(  \mathbb{W}(v)+v^{2}\right)  ],
\end{align}
where avar stands for asymptotic variance and var for variance.

In order to determine the finite sample behavior of the isotonic M-estimators
we have performed a Monte Carlo study. \ We took i.i.d.\ samples from the
model (\ref{eqn:model}) with trend term $\mu(t)=10+5t^{2}$ and where the
distribution $G$ is N(0,1) and Student with three degrees of freedom. The
values $\{t_{i}=i/(n+1),1\leq i\leq n\}$ corresponds to a uniform limiting
distribution $H(t)=t$ for $0<t<1$.

\ We estimated $\mu(t_{0})$ at $t_{0}=1/2$, the true value of which is
$\mu(t_{0})=11.25$ using three isotonic estimators: the L$_{2}$ isotonic
estimator, the L$_{1}$ isotonic estimator and the same isotonic M-estimator
that was used in the examples. We performed $N=500$ replicates at two sample
sizes, $n=100$ and $500$. Dykstra and Carolan (1998) have established that the
variance of the random variable \textquotedblleft$\text{slogcm}\left(
\mathbb{W}(v)+v^{2}\right)  $\textquotedblright\ is approximately 1.04. Using
this value, we present in Table 1 sample mean square errors (MSE)\ times
$n^{2/3}$ as well as the corresponding asymptotic variances.%

\[%
\begin{tabular}
[c]{ccccccc|}\hline
Estimator & \multicolumn{2}{c}{n = 100} & \multicolumn{2}{c}{n=500} &
\multicolumn{2}{c|}{avar}\\\cline{2-7}
& Normal & Student$_{3}$ & Normal & Student$_{3}$ & Normal & Student$_{3}%
$\\\hline
L$_{2}$ & 1.93 & 3.78 & 1.85 & 3.65 & 1.92 & 3.98\\
L$_{1}$ & 2.38 & 2.89 & 2.67 & 2.76 & 2.59 & 2.89\\
M & 2.04 & 2.86 & 2.11 & 2.51 & 2.06 & 2.53\\\hline
\end{tabular}
\ \ \
\]

\begin{center}
Table 1. Sample MSE and avar for Isotonic Regression Estimators.
\end{center}

We note that for \ both distributions, the empirical MSEs \ for $n=500$ are
close to the avar values. \ We also see that under both distributions \ the
M-estimator is more efficient that the L$_{1}$ one, that the M- estimator is
more efficient than the L$_{1}$ one \ for both distributions and that the
L$_{1}$ estimator is slightly less efficient than the L$_{2}$ estimator for
the normal case but much more efficient for the Student distribution. \ In
summary, \ the isotonic M-estimate seems to have a good behavior under both distributions.

\section{Appendix\label{Appendix}}

\subsection{Proof of Lemma 1}

Without loss of generality \ we can assume that $\sigma_{0}=1.$ Given $c>0$,
for sufficiently large $n$ there exist positive numbers $\beta_{l}(n)$ and
$\beta_{u}(n)$ for which
\[
H(t_{0})-H(t_{0}-\beta_{l}(n))=H(t_{0}+\beta_{u}(n))-H(t_{0})=cn^{-1/3}.
\]
As in Wright (1981), we first argue that
\[
\mathrm{P}[\hat{\mu}_{n}(t_{0})\neq\hat{\mu}_{n}^{\ast}(t_{0})]\leq
\mathrm{P}(\Omega_{1n})+\mathrm{P}(\Omega_{2n}),
\]
where
\begin{align}
\Omega_{1n}  &  =\left\{  \min_{v\geq t_{0}}\hat{\mu}_{n}(t_{0}-\beta
_{l}(n),v]\right.  <\left.  \max_{u\leq t_{0}-\alpha_{l}(n)}\hat{\mu}%
_{n}[u,t_{0}-\beta_{l}(n)]\right\}  ,\label{omega1}\\
\Omega_{2n}  &  =\left\{  \max_{u\leq t_{0}}\hat{\mu}_{n}[u,t_{0}+\beta
_{u}(n))\right.  >\left.  \min_{v\geq t_{0}+\alpha_{u}(n)}\hat{\mu}_{n}%
[t_{0}+\beta_{u}(n),v]\right\}  . \label{eqn:omega2}%
\end{align}
To see this, note that the complement of $\Omega_{2n}$ is the set in which,
for all $u\leq t_{0}$ and all $v\geq t_{0}+\beta_{u}(n)$ we have that
$\left\{  \hat{\mu}_{n}[u,t_{0}+\beta_{u}(n))\leq\hat{\mu}_{n}[t_{0}+\beta
_{u}(n),v]\right\}  $. Since $\psi$ is non-decreasing we can write
\[
\hat{\mu}_{n}[u,t_{0}+\beta_{u}(n))\leq\hat{\mu}_{n}[u,v].
\]
\ This in turn entails that in $\Omega_{2n}^{c}$
\[
\mu_{n}^{\ast}(t_{0})=\max_{u\leq t_{0}}\min_{t_{0}\leq v<t_{0}+\alpha_{u}%
(n)}\hat{\mu}_{n}[u,v].
\]
Using the fact that the maximum and the minimum may be reversed in computing
these estimators (e.g.\ Robertson and Waltman, 1968) and a similar argument
for $\Omega_{1n}$ in equation (\ref{omega1}) one can show that
\[
\mathrm{P}\{\Omega_{1n}^{c}\cap\Omega_{2n}^{c}\}\leq\mathrm{P}\{\mu_{n}^{\ast
}(t_{0})=\hat{\mu}_{n}(t_{0})\}.
\]
So we need to prove that
\[
\lim_{c\rightarrow\infty}\limsup_{n\rightarrow\infty}\mathrm{P}(\Omega
_{1n})=\lim_{c\rightarrow\infty}\limsup_{n\rightarrow\infty}\mathrm{P}%
(\Omega_{2n})=0.
\]
We will prove $\lim_{c\rightarrow\infty}\limsup_{n\rightarrow\infty}%
\mathrm{P}(\Omega_{1n})=0.$\ The result for $\Omega_{2n}$ can be obtained in a
similar manner.\ 

Let
\begin{align}
\Lambda_{1n}  &  =\left\{  \min_{v\geq t_{0}}\hat{\mu}_{n}(t_{0}-\beta
_{l}(n),v]<\mu(t_{0}-\beta_{l}(n))\right\}  ,\label{eqn:bound1}\\
\Lambda_{2n}  &  =\left\{  \max_{u\leq t_{0}-\alpha_{l}(n)}\hat{\mu}%
_{n}[u,t_{0}-\beta_{l}(n)]>\mu(t_{0}-\beta_{l}(n))\right\}  .
\label{eqn:bound2}%
\end{align}
\ Since
\[
\mathrm{P}(\Omega_{1n})\leq\mathrm{P}(\Lambda_{1n})+\mathrm{P}(\Lambda_{2n}),
\]
it will be enough to prove that%
\begin{equation}
\lim_{c\rightarrow\infty}\limsup_{n\rightarrow\infty}\mathrm{P}(\Lambda
_{in})=0,i=1,2. \label{lama1}%
\end{equation}
Since the proofs of (\ref{lama1}) for $i=1$ and $2$ \ are similar,
(\ref{lama1}) will be only proved for $i=1.$ By the fundamental identity
(\ref{eqn:equivs}) we have
\begin{equation}
\Lambda_{1n}=\left\{  \min_{v\geq t_{0}}S_{n}\left(  t_{0}-\beta_{l}%
(n),v,\mu(t_{0}-\beta_{l}(n))\right)  <0\right\}  . \label{eqn:l}%
\end{equation}
In the sequel in order to simplify notation we will omit the subindex $n$
writing $x_{n,j}=x_{j}$, $t_{n,j}=t_{j}$ and $u_{n,j}=u_{j}$ making it
explicit only when there is a risk of confusion. We can write%
\begin{align*}
S_{n}\left(  t_{0}-\beta_{l}(n),v,\mu(t_{0}-\beta_{l}(n))\right)   &
=\sum\limits_{j\in C(t_{0}-\beta_{l}(n),v)}\psi\left(  x_{j}-\mu(t_{0}%
-\beta_{l}(n)\right) \\
&  =\sum\limits_{j\in C(t_{0}-\beta_{l}(n),v)}\psi\left(  x_{j}-\mu(t_{j}%
)+\mu(t_{j})-\mu(t_{0}-\beta_{l}(n)\right) \\
&  =\sum\limits_{j\in C(t_{0}-\beta_{l}(n),v)}\psi\left(  u_{j}+(\mu
(t_{j})-\mu(t_{0}-\beta_{l}(n))\right)  ,
\end{align*}
and by a Taylor expansion we get%
\[
S_{n}\left(  t_{0}-\beta_{l}(n),v,\mu(t_{0}-\beta_{l}(n))\right)
=\sum\limits_{j\in C(t_{0}-\beta_{l}(n),v)}\left[  \psi(u_{j})+\psi^{\prime
}(u_{j}+a_{j}^{\ast})(\mu(t_{j})-\mu(t_{0}-\beta_{l}(n))\right]  ,
\]
where $0\leq a_{j}^{\star}\leq\mu(t_{j})-\mu(t_{0}-\beta_{l}(n)$. Put
$\tau=\sup\psi^{\prime}$, then
\[
S_{n}\left(  t_{0}-\beta_{l}(n),v,\mu(t_{0}-\beta_{l}(n))\right)  \leq
\sum\limits_{j\in C(t_{0}-\beta_{l}(n),v)}\psi(u_{j})+\tau\sum\limits_{j\in
C(t_{0}-\beta_{l}(n),v)}(\mu(t_{j})-\mu(t_{0}-\beta_{l}(n)).
\]
Thus, since $\mu(t)$ is increasing we get
\begin{equation}
\min_{v\geq t_{0}}S_{n}\left(  t_{0}-\beta_{l}(n),v,\mu(t_{0}-\beta
_{l}(n))\right)  \leq\min_{v\geq t_{0}}\sum\limits_{j\in C(t_{0}-\beta
_{l}(n),v)}\psi(u_{j})+\tau\sum\limits_{j\in C(t_{0}-\beta_{l}(n),t_{0})}%
(\mu(t_{j})-\mu(t_{0}-\beta_{l}(n)). \label{203}%
\end{equation}
Put $n_{l}(v):=\#\{j:t_{0}-\beta_{l}(n)\leq t_{nj}\leq v\}$. As $n_{l}(v)\geq
n_{l}(t_{0})$, we obtain
\begin{align}
&  \min_{v\geq t_{0}}\frac{1}{n_{l}(v)}S_{n}\left(  t_{0}-\beta_{l}%
(n),v,\mu(t_{0}-\beta_{l}(n))\right) \label{eqn:3}\\
&  \leq\min_{v\geq t_{0}}\frac{1}{n_{l}(v)}\sum\limits_{j\in C(t_{0}-\beta
_{l}(n),v)}\psi(u_{j})+\frac{\tau}{n_{l}(t_{0})}\sum\limits_{j\in
C(t_{0}-\beta_{l}(n),t_{0})}(\mu(t_{j})-\mu(t_{0}-\beta_{l}(n)).
\end{align}
Therefore the event $\Lambda_{1n}$ defined in (\ref{eqn:l}) is included in the
event $\Delta_{n}$ defined by
\[
\Delta_{n}=\left\{  \max_{v\geq t_{0}}\frac{1}{n_{l}(v)}\sum_{j\in
C(t_{0}-\beta_{l}(n),v)}-\psi(u_{j})>\frac{\tau}{n_{l}(t_{0})}\sum_{j\in
C(t_{0}-\beta_{l}(n),t_{0})}(\mu(t_{j})-\mu(t_{0}-\beta_{l}(n))\right\}  .
\]
The equation above can be rewritten in terms of integrals with respect to the
empirical distribution of the $t$'s as
\begin{equation}
\max_{v\geq t_{0}}\frac{1}{n_{l}(v)}\sum_{j\in C(t_{0}-\beta_{l}(n),v)}%
-\psi(u_{j})>\tau\int_{t_{0}-\beta_{l}(n)}^{t_{0}}|\mu(s)-\mu(t_{0}-\beta
_{l}(n))|dH_{n}(s). \label{eqn:boundd}%
\end{equation}
Since $u_{i},\ldots,u_{n}$ are i.i.d., relabelling the $u_{j}^{\text{ }}$'s on
the left hand side we get that
\begin{equation}
\mathrm{P}(\Lambda_{1n})\leq\mathrm{P}(\Delta_{n}^{\ast}), \label{delta1<}%
\end{equation}
where
\begin{equation}
\Delta_{n}^{\ast}=\left\{  \max\limits_{n_{l}(t_{0})\leq k\leq n}\frac{1}%
{k}\sum_{n_{l}(t_{0})\leq j\leq k}-\psi(u_{j})>2\tau\int_{t_{0}-\beta_{l}%
(n)}^{t_{0}}|\mu(s)-\mu(t_{0}-\beta_{l}(n))|dH_{n}(s)\right\}  .
\label{eqn:eqsum}%
\end{equation}
\ Adding and subtracting $dH(s)$ we can write
\begin{align}
\int_{t_{0}-\beta_{l}(n)}^{t_{0}}[\mu(s)-\mu(t_{0}-\beta_{l}(n))]dH_{n}(s)  &
=\int_{t_{0}-\beta_{l}(n)}^{t_{0}}[\mu(s)-\mu(t_{0}-\beta_{l}%
(n))]dH(s)\nonumber\\
&  +\int_{t_{0}-\beta_{l}(n)}^{t_{0}}[\mu(s)-\mu(t_{0}-\beta_{l}%
(n))]d(H_{n}(s)-H(s)). \label{decomp}%
\end{align}
Using (\ref{condH}), for $n$ large enough, the second term in the above
equation is bounded by
\begin{align*}
\left\vert \int_{t_{0}-\beta_{l}(n)}^{t_{0}}[\mu(s)-\mu(t_{0}-\beta
_{l}(n))]dH_{n}(s)-H(s))\right\vert  &  \leq2\left(  \mu(t_{0})-\mu
(t_{0}-\beta_{l}(n)\right)  \sup_{t}|H_{n}(t)-H(t)|\\
&  \leq2\mu^{\prime}(t_{0})\beta_{l}(n)n^{-1/3}o(1),
\end{align*}
and since by the inverse function theorem $\beta_{l}(n)=c[H^{\prime}%
(t_{0})]^{-1}n^{-1/3}[1+o(1)]$, we obtain that for some constant $A$ which
does not depend on $c$ we can write
\begin{equation}
\left\vert \int_{t_{0}-\beta_{l}(n)}^{t_{0}}[\mu(s)-\mu(t_{0}-\beta
_{l}(n))]d(H_{n}(s)-H(s))\right\vert \leq Acn^{-2/3}o(1). \label{intbrav1}%
\end{equation}
Consider now the first term in the right hand side of Equation (\ref{decomp}).
Using (\ref{condH}) we have%
\begin{align*}
\int\limits_{(t_{0}-\beta_{l}(n),t_{0}]}  &  [\mu(s)-\mu(t_{0}-\beta
_{l}(n))]dH(s)\\
&  =\int_{t_{0}-\beta_{l}(n)}^{t_{0}}[\mu(t_{0})-\mu(t_{0}-\beta
_{l}(n))]dH(s)-\int_{t_{0}-\beta_{l}(n)}^{t_{0}}[\mu(t_{0})-\mu(s)]dH(s)\\
&  =[\mu(t_{0})-\mu(t_{0}-\beta_{l}(n))][H(t_{0})-H(t_{0}-\beta_{l}%
(n))]-\int_{t_{0}-\beta_{l}(n)}^{t_{0}}[\mu(t_{0})-\mu(s)]dH(s)\\
&  \leq\left(  \frac{\mu(t_{0})-\mu(t_{0}-\beta_{l}(n))}{\beta_{l}(n)}\right)
\left(  \frac{H(t_{0})-H(t_{0}-\beta_{l}(n))}{\beta_{l}(n)}]\right)  \beta
_{l}(n)^{2}\\
&  =\mu^{\prime}(t_{0})[1+o(1)]H^{\prime}(t_{0})[1+o(1)]\beta_{l}(n)^{2}\\
&  =\mu^{\prime}(t_{0})[H^{\prime}(t_{0})]^{-1}c^{2}n^{-2/3}[1+o(1)].
\end{align*}
Therefore
\begin{align*}
\int_{t_{0}-\beta_{l}(n)}^{t_{0}}  &  [\mu(s)-\mu(t_{0}-\beta_{l}%
(n))]dH_{n}(s)\\
&  \leq\mu^{\prime}(t_{0})c^{2}[H^{\prime}(t_{0})]^{-1}n^{-2/3}(1+o(1))+2\mu
^{\prime}(t_{0})c[H^{\prime}(t_{0})]^{-1}n^{-2/3}o(1)\\
&  \leq\mu^{\prime}(t_{0})c^{\star}[H^{\prime}(t_{0})]^{-1}n^{-1/3}\left\{
n^{-1/3}[1+o(1)]+2n^{-1/3}o(1)\right\}
\end{align*}
with $c^{\star}=\max(c,c^{2})$. Then, for some constant $B$ which does not
depend on $c$ we can write
\begin{equation}
\int_{t_{0}-\beta_{l}(n)}^{t_{0}}[\mu(s)-\mu(t_{0}-\beta_{l}(n))]dH_{n}(s)\leq
Bc^{\star}n^{-1/3}. \label{intB}%
\end{equation}
\ From (\ref{eqn:l}), (\ref{eqn:boundd}), (\ref{delta1<}), (\ref{eqn:eqsum}),
(\ref{decomp}), (\ref{intbrav1}) and (\ref{intB}) we derive that there exists
a constant $D$ independent of $c$ such that for $n$ large enough and $c>1$
\begin{equation}
\mathrm{P}(\Lambda_{1n})\leq\text{P}\left\{  \max\limits_{n_{l}(t_{0})\leq
k\leq n}\frac{1}{k}\sum_{1\leq j\leq k}-\psi(u_{j})>Dc^{2}n^{-1/3}\ \right\}
. \label{eqn:tohr}%
\end{equation}
At this point, we use the H\`{a}jek-Renyi Maximal Inequality (e.g., Shorack,
2000) which asserts that for a sequence $y_{1},\ldots,y_{n}$ of independent
random variables with mean $0$ and finite variances and for a positive
non-decreasing real sequence $\{b_{k},k\in N\}$,
\begin{equation}
\text{P}\left\{  \max\limits_{m\leq k\leq n}\left\vert \frac{\sum_{j=1}%
^{k}y_{j}}{b_{k}}\right\vert \geq\lambda\right\}  \leq\frac{1}{\lambda^{2}%
}\left\{  \sum_{k=1}^{m}\frac{\text{E}(y_{k}^{2})}{b_{m}^{2}}+\sum_{k=m+1}%
^{n}\frac{\text{E}(y_{k}^{2})}{b_{k}^{2}}\right\}  . \label{eqn:HR}%
\end{equation}
Using this inequality from (\ref{eqn:tohr}) we get that
\begin{equation}
\text{P}(\Lambda_{1n})\leq\text{P}\left\{  \max\limits_{n_{l}(t_{0})\leq k\leq
n}\sum_{1\leq j\leq k}\frac{-\psi(u_{j})}{k}>Dc^{2}n^{-1/3}\right\}  \leq
\frac{\text{\textrm{E}$_{G}$(}\psi^{2}(u))\left(  \frac{1}{n_{l}^{2}(t_{0}%
)}+\sum_{k=n_{l}(t_{0})}^{n}\frac{1}{k^{2}}\right)  }{D^{2}c^{4}n^{-2/3}}.
\label{file1}%
\end{equation}
\ Approximating the Riemann sum we obtain
\begin{equation}
\sum_{k=n_{l}(t_{0})}^{n}k^{-2}\leq\frac{1}{n_{l}(t_{0})} \label{file2}%
\end{equation}
and since by (\ref{condH}) $n_{l}(t_{0})=cn^{2/3}(1+o(1))$, for $n$ large
enough we have
\begin{equation}
n_{l}(t_{0})^{-1}\leq2c^{-1}n^{-2/3}. \label{file3}%
\end{equation}
From (\ref{file1}), (\ref{file2}) and (\ref{file3}) we derive that for $\ n$
large enough
\begin{align*}
\text{P}(\Lambda_{1n})  &  \leq\frac{2\text{\textrm{E}$_{G}$}(\psi^{2}%
(u))}{D^{2}c^{4}n^{-2/3}n_{l}(t_{0})}\\
&  \leq\frac{4\text{\textrm{E}$_{G}$}(\psi^{2}(u))}{D^{2}c^{5}}.
\end{align*}
Then the Lemma follows immediately.

\subsection{Proof of Theorem 1}

Without loss of generality we can assume that $\sigma_{0}=1.$Since $\alpha
_{l}(n)=\alpha_{u}(n)=2c[H^{\prime}(t_{0})]^{-1}n^{-1/3}[1+o(1)]$, and $c$ is
arbitrary, we will consider the localized estimator
\begin{align}
\hat{\mu}_{n}^{c}(t_{0})  &  =\max_{t_{0}-cn^{-1/3}<u\leq t_{0}}\min
_{t_{0}\leq v<t_{0}+cn^{-1/3}}\hat{\mu}_{n}(u,v)\nonumber\\
&  =\max_{u\leq t_{0}}\min_{v\geq t_{0}}\hat{\mu}_{n}^{c}(u,v),
\label{eqn:modest2}%
\end{align}
where $\hat{\mu}_{n}^{c}(u,v)$ is defined as the root of
\begin{equation}
S_{n}^{c}(u,v,\mu)=\sum_{j\in D(u,v)}\psi(x_{j}-\mu) \label{eqn:partialsumsb}%
\end{equation}
over $D(u,v)=\{j:1\leq j\leq n;t_{0}-cn^{-1/3}<u\leq t_{j}\leq v<t_{0}%
+cn^{-1/3}\}$. Note that the localized estimator depends on the $t_{j}$'s that
lie on a neighborhood about $t_{0}$ which shrinks at a rate $n^{-1/3}$. To
proceed with the development of the asymptotic distribution let now
$w_{j}=n^{1/3}(t_{j}-t_{0})$, $r=n^{1/3}(u-t_{0})$ and $s=n^{1/3}(v-t_{0})$.
With this notation, $\hat{\mu}_{n}^{c}(u,v)$ is a root of the partial sums in
the parametrization
\begin{equation}
\dot{S}_{n}^{c}(r,s,\mu)=\sum_{j\in B(r,s)}\psi(x_{j}-\mu)=0,
\label{eqn:partialsums2}%
\end{equation}
where $B(r,s)=\{j:1\leq j\leq n;r\leq w_{j}\leq s;r,s\in\lbrack-c,c]\}$. So
that the relabelling implies $\hat{\mu}_{n}^{c}(u,v)\equiv\dot{\mu}_{n}%
^{c}(r,s)$. Consequently,
\[
\hat{\mu}_{n}^{c}(t_{0})=\max_{r\leq0}\min_{v\geq0}\dot{\mu}_{n}^{c}(r,s).
\]
Now a Taylor expansion of $\mu(t_{j})$ around $t_{0}$ for any $j\in B(r,s)$
gives
\begin{align*}
\mu(t_{j})  &  =\mu(t_{0})+\mu^{\prime}(t_{0})(t_{j}-t_{0})+o(|t_{j}-t_{0}|)\\
&  =\mu(t_{0})+\mu^{\prime}(t_{0})n^{-1/3}w_{i}+o_{j}(n^{-1/3})
\end{align*}
which entails that
\[
x_{j}=\mu(t_{0})+\mu^{\prime}(t_{0})n^{-1/3}w_{j}+u_{j}+o_{j}(n^{-1/3}).
\]
\ Using the equivariance of M-estimators, the monotonicity of $\psi$ and the
fact that $\psi^{^{\prime\prime}}$ is bounded, it can be proved that
\[
\dot{\mu}_{n}^{c}(r,s)=\mu(t_{0})+\tilde{\mu}_{n}^{c}(r,s)+o_{rs}(n^{-1/3}),
\]
where $\tilde{\mu}_{n}^{c}(r,s)$ solves
\begin{equation}
\tilde{S}_{n}^{c}(r,s,\mu)=\sum_{j\in B(r,s)}\psi(n^{-1/3}\mu^{\prime}%
(t_{0})w_{j}+u_{j}-\mu)=0 \label{murs}%
\end{equation}
and%
\[
\left\vert o_{rs}(n^{-1/3})\right\vert \leq K_{1}c^{2}n^{-2/3}.
\]
Thus, using that the $w_{j}$ are bounded over $-c\leq r\leq w_{i}\leq s\leq c$
we have
\[
\hat{\mu}_{n}^{c}(t_{0})=\mu(t_{0})+\max_{-c\leq r\leq0}\min_{0\leq v\leq
c}\tilde{\mu}_{n}^{c}(r,s)+o(n^{-1/3}).
\]
This entails that
\[
n^{1/3}[\hat{\mu}_{n}^{c}(t_{0})-\mu(t_{0})]=\max_{r\leq0}\min_{v\geq0}%
n^{1/3}\tilde{\mu}_{n}^{c}(r,s)+o_{rs}^{\ast}(1),
\]
where%
\[
\left\vert o_{rs}^{{\ast}}(n^{-1/3})\right\vert \leq K_{2}c^{2}.
\]
Then, we \ only need to obtain the asymptotic distribution of
\[
\Delta_{n,c}=n^{1/3}\max_{r\leq0}\min_{s\geq0}\tilde{\mu}_{n}^{c}(r,s).
\]

Let $\tilde{\mu}_{n}^{c\ast}(r,s)$ be the solution of
\[
\sum_{j\in B(r,s)}\psi(u_{j}-\mu)=0.
\]
Since $|n^{-1/3}\mu^{\prime}(t_{0})w_{j}|\leq n^{-1/3}\mu^{\prime}(t_{0})c$ we
have that
\begin{equation}
\tilde{\mu}_{n}^{c}(r,s)=\tilde{\mu}_{n}^{c\ast}(r,s)+d_{nrs}, \label{dnrs1}%
\end{equation}
where%
\begin{equation}
|d_{nrs}|\leq K_{3}cn^{-1/3}. \label{dnrs2}%
\end{equation}
We will approximate now $n_{rs}$ as follows$\ $%
\begin{align}
\frac{n_{rs}}{n}  &  =\frac{1}{n}\#\{1\leq j\leq n:r\leq0\leq w_{j}\leq
s\}\nonumber\\
&  =\frac{1}{n}\#\{1\leq j\leq n:t_{0}-rn^{-1/3}\leq t_{0}\leq t_{j}\leq
t_{0}+sn^{-1/3}\}\nonumber\\
&  =H_{n}(t_{0}+sn^{-1/3})-H_{n}(t_{0}-rn^{-1/3})\nonumber\\
&  =[H_{n}(t_{0}+sn^{-1/3})-H(t_{0}+sn^{-1/3})]\nonumber\\
&  +[H(t_{0}+sn^{-1/3})-H(t_{0}-rn^{-1/3})]-[H_{n}(t_{0})-H(t_{0})]\nonumber\\
&  =H^{\prime}(t_{0})(s-r)n^{-1/3}+o(n^{-1/3})\nonumber\\
&  =n^{-1/3}H^{\prime}(t_{0})(s-r)[1+o(1)], \label{nrsfor}%
\end{align}
and therefore%
\begin{equation}
n_{rs}=n^{2/3}H^{\prime}(t_{0})(s-r)[1+o(1)], \label{nrsfor2}%
\end{equation}%
\begin{equation}
n_{rs}^{1/2}=n^{1/3}H^{\prime}(t_{0})^{1/2}(s-r)^{1/2}(1+o(1)) \label{nrsfor3}%
\end{equation}
and%
\begin{equation}
\frac{n^{1/2}}{n_{rs}}=\frac{1}{n_{rs}^{1/3}H^{\prime}(t_{0})^{1/2}%
(s-r)^{1/2}(1+o(1))}. \label{nrsfor4}%
\end{equation}
Then, taking $n_{rs}\rightarrow\infty$ and applying the law of large numbers
is easy to show that $\tilde{\mu}_{n}^{c\ast}(r,s)\rightarrow\mu_{0}$ a.s. and
therefore by (\ref{dnrs1}) and (\ref{dnrs2}) $\tilde{\mu}_{n}^{c\ast
}(r,s)\rightarrow_{p}\mu_{0}$ too. Since $\tilde{\mu}_{n}^{c}(r,s)$ satisfies
(\ref{murs}), by a Taylor expansion of $\tilde{S}_{n}^{c}(r,s)$ we get
\begin{align*}
&  \sum_{B(r,s)}\psi(u_{j})-\sum_{B(r,s)}\psi^{\prime}(u_{j})(\tilde{\mu}%
_{n}^{c}(r,s)-n^{-1/3}\mu^{\prime}(t_{0})w_{j})\\
&  +\sum_{B(r,s)}\psi^{\prime\prime}(\varepsilon_{j}^{\ast})(\tilde{\mu}%
_{n}^{c}(r,s)-n^{-1/3}\mu^{\prime}(t_{0})w_{j})^{2}\\
&  =0.
\end{align*}
\bigskip From here we obtain
\begin{align*}
&  \tilde{\mu}_{n}^{c}(r,s)\\
&  =\frac{\sum_{j\in B(r,s)}\psi(u_{j})+\mu^{\prime}(t_{0})n^{-1/3}\sum_{j\in
B(r,s)}w_{j}\psi^{\prime}(u_{j})+n^{-2/3}\mu^{\prime}(t_{0})^{2}\sum_{j\in
B(r,s)}w_{j}^{2}\psi^{\prime\prime}(\varepsilon_{j}^{\ast})}{\sum_{j\in
B(r,s)}\psi^{\prime}(u_{j})-\tilde{\mu}_{n}^{c}(r,s)\sum_{j\in B(r,s)}%
\psi^{\prime\prime}(\varepsilon_{j}^{\ast})^{2}-2n^{-1/3}\mu^{\prime}%
(t_{0})\sum_{j\in B(r,s)}w_{j}\psi^{\prime\prime}(\varepsilon_{j}^{\ast})}%
\end{align*}
and then%
\begin{align}
&  n^{1/3}\tilde{\mu}_{n}^{c}(r,s)\nonumber\\
&  =\frac{\frac{n^{1/3}}{n^{2/3}}\ \sum_{j\in B(r,s)}\psi(u_{j})+\mu^{\prime
}(t_{0})\frac{1}{n^{2/3}}\sum_{j\in B(r,s)}w_{j}\psi^{\prime}(u_{j}%
)+n^{-1/3}\mu^{\prime}(t_{0})^{2}\frac{1}{n^{2/3}}\sum_{j\in B(r,s)}w_{j}%
^{2}\psi^{\prime\prime}(\varepsilon_{j}^{\ast})}{\frac{1}{n^{2/3}}\sum_{j\in
B(r,s)}\psi^{\prime}(u_{j})-\tilde{\mu}_{n}^{c}(r,s)\frac{1}{n^{2/3}}%
\sum_{j\in B(r,s)}\psi^{\prime\prime}(\varepsilon_{j}^{\ast})^{2}-2n^{-1/3}%
\mu^{\prime}(t_{0})\frac{1}{n^{2/3}}\sum_{j\in B(r,s)}w_{j}\psi^{\prime\prime
}(\varepsilon_{j}^{\ast})}. \label{largui1}%
\end{align}
By (\ref{murs}), the Law of the Large Numbers, $|w_{j}|\leq c$ and
$\psi^{\prime\prime}$ bounded we have%
\[
n^{-1/3}\mu^{\prime}(t_{0})^{2}\frac{1}{n^{2/3}}\sum_{j\in B(r,s)}w_{j}%
^{2}\psi^{\prime\prime}(\varepsilon_{j}^{\ast})\rightarrow0,
\]%
\[
2n^{-1/3}\mu^{\prime}(t_{0})\frac{1}{n^{2/3}}\sum_{j\in B(r,s)}w_{j}%
\psi^{\prime\prime}(\varepsilon_{j}^{\ast})\rightarrow0,
\]%
\[
\tilde{\mu}_{n}^{c}(r,s)\frac{1}{n^{2/3}}\sum_{j\in B(r,s)}\psi^{\prime\prime
}(\varepsilon_{j}^{\ast})^{2}\rightarrow0
\]
and
\[
\frac{1}{n^{2/3}}\sum_{j\in B(r,s)}\psi^{\prime}(u_{j})\rightarrow
(s-r)H^{\prime}(t_{0})\mathrm{\mathrm{E}}_{G}\mathrm{(}\psi^{\prime}(u))\text{
a.s..}%
\]
Then, (\ref{largui1}) entails%
\begin{equation}
(s-r)\mathrm{\mathrm{E}_{G}}(\psi^{\prime}(u))H^{\prime}(t_{0})\ n^{1/3}%
\tilde{\mu}_{n}^{c}(r,s)=\frac{1}{n^{1/3}}\sum_{j\in B(r,s)}\psi(u_{j}%
)+\mu^{\prime}(t_{0})\frac{1}{n^{2/3}}\sum_{j\in B(r,s)}w_{j}\psi^{\prime
}(u_{j})+o_{rs}(1). \label{nosac}%
\end{equation}
Let
\begin{equation}
B_{n}(s)=%
\begin{cases}
\dfrac{\mu^{\prime}(t_{0})}{n^{1/3}\mathrm{\mathrm{E}_{G}}(\psi^{2}%
(u))^{1/2}H^{\prime}(t_{0})^{1/2}}\sum\limits_{j\in B(0,s)}\psi(u_{j}) &
\text{if }s>0\\
\dfrac{\mu^{\prime}(t_{0})}{n^{1/3}\mathrm{\mathrm{E}_{G}}(\psi^{2}%
(u))^{1/2}H^{\prime}(t_{0})^{1/2}}\sum\limits_{j\in B(s,0)}-\psi(u_{j}) &
\text{if }s<0,
\end{cases}
. \label{Bns}%
\end{equation}
By (\ref{nrsfor3}) and the Central Limit Theorem we have that for any set of
finite numbers $s_{1},s_{2},...,s_{r}$,$-c\leq s_{i}\leq c,$ the random vector
$(B_{n,s_{1}},...,B_{n,s_{r}})$ converges in distribution to N(0, $\Sigma)$
where $\Sigma=(\sigma_{ij})$ with
\[
\sigma_{ij}=%
\begin{cases}
s_{i}\wedge s_{j} & \text{if }s_{i}\geq0,s_{j}\geq0\\
-s_{i}\wedge-s_{j} & \text{if }s_{i}\leq0,s_{j}\leq0\\
0 & \text{if }s_{i}\geq0,s_{j}\leq0
\end{cases}
.
\]
Moreover, using standard arguments, it can be proved that $B_{n}(s)$ is tight.
Then, we have
\begin{equation}
B_{n}(s)\overset{\mathcal{D}}{\Rightarrow}B(s), \label{bnsac1}%
\end{equation}
where $B$ is a two sided Brownian motion

As for the second term in the right hand side of (\ref{nosac}) define
\begin{equation}
\Lambda_{n}(s)=%
\begin{cases}
\dfrac{1}{n^{2/3}}\sum\limits_{0\leq w_{j}\leq s}\psi^{\prime}(u_{j})w_{j} &
\text{if }s>0\\
\dfrac{1}{n^{2/3}}\sum\limits_{s\leq w_{j}\leq0}-\psi^{\prime}(u_{j})w_{j} &
\text{if }s<0
\end{cases}
. \label{bnsac2}%
\end{equation}
For $s>0$ we can write%
\begin{equation}
\Lambda_{n}(s)=\frac{1}{n}\sum\limits_{j=1}^{n}n^{2/3}\psi^{\prime}%
(u_{j})(t_{j}-t_{0})1(t_{0}\leq t_{j}\leq t_{0}+sn^{-1/3}) \label{Lan0}%
\end{equation}
and then
\begin{equation}
\text{E}(\Lambda_{n}(s))=\text{\textrm{E}$_{G}$}(\psi^{\prime}(u_{j}%
))\int_{t_{0}}^{t_{0}+sn^{-1/3}}n^{2/3}(t-t_{0})dH_{n}\ \label{Lan1}%
\end{equation}
Integrating by parts we get
\begin{equation}
\int_{t_{0}}^{t_{0}+sn^{-1/3}}n^{2/3}(t-t_{0})dH_{n}\ =n^{1/3}s^{2}H_{n}%
(t_{0}+sn^{-1/3})-\int_{t_{0}}^{t_{0}+sn^{-1/3}}n^{2/3}H_{n}(t)dt \label{Lan2}%
\end{equation}
and by (\ref{condH}) we have
\begin{equation}
n^{1/3}s^{2}H_{n}(t_{0}+sn^{-1/3})=n^{1/3}s^{2}H(t_{0}+sn^{-1/3})+o(1).
\label{Lan3}%
\end{equation}
We can write
\[
\int_{t_{0}}^{t_{0}+sn^{-1/3}}n^{2/3}H_{n}(t)dt=\int_{t_{0}}^{t_{0}+sn^{-1/3}%
}n^{2/3}H(t)dt+\int_{t_{0}}^{t_{0}+sn^{-1/3}}n^{2/3}(H_{n}(t)-H(t))dt,
\]
and by(\ref{condH}) we get
\begin{equation}
\left\vert \int_{t_{0}}^{t_{0}+sn^{-1/3}}n^{2/3}(H_{n}(t)-H(t))dt\right\vert
\leq sn^{-1/3}n^{1/3}\sup_{t}n^{1/3}\left\vert H_{n}(t)-H(t)\right\vert =o(1).
\label{Lan4}%
\end{equation}
Therefore by (\ref{Lan2}), (\ref{Lan3}) and (\ref{Lan4}) we get%
\begin{align*}
\int_{t_{0}}^{t_{0}+sn^{-1/3}}n^{2/3}(t-t_{0})dH_{n}  &  =n^{1/3}s^{2}%
H_{n}(t_{0}+sn^{-1/3})-\int_{t_{0}}^{t_{0}+sn^{-1/3}}n^{2/3}H_{n}(t)dt+o(1)\\
&  =\int_{t_{0}}^{t_{0}+sn^{-1/3}}n^{2/3}(t-t_{0})dH+o(1)\\
&  =\int_{t_{0}}^{t_{0}+sn^{-1/3}}n^{2/3}(t-t_{0})H^{\prime}(t)dt+o(1)\\
&  =H^{\prime}(t_{0})\int_{t_{0}}^{t_{0}+sn^{-1/3}}n^{2/3}(t-t_{0})dt+o(1)\\
&  =H^{\prime}(t_{0})\frac{s^{2}}{2}+o(1),
\end{align*}
and for (\ref{Lan1}) we get that for $s>0$
\begin{equation}
\text{E}(\Lambda_{n}(s))=\text{\textrm{E}$_{G}$}(\psi^{\prime}(u_{j}%
))H^{\prime}(t_{0})\frac{s^{2}}{2}+o(1). \label{Lan5}%
\end{equation}

Now we compute the variance of $\Lambda_{n}(s).$ From (\ref{Lan0}) we have%
\[
\Lambda_{n}(s)=\frac{1}{n}\sum\limits_{j=1}^{n}n^{2/3}\psi^{\prime}%
(u_{j})(t_{j}-t_{0})1\left(  t_{0}\leq t_{j}\leq t_{0}+sn^{-1/3}\right)
\]

\begin{align*}
\text{var}(\Lambda_{n}(s))  &  =\frac{\text{var}(\psi^{\prime}(u))}{n}%
\sum\limits_{j=1}^{n}n^{1/3}(t_{j}-t_{0})^{2}1(t_{0}\leq t_{j}\leq
t_{0}+sn^{-1/3})\\
&  =n^{1/3}\text{var}(\psi^{\prime}(u))\int_{t_{0}}^{t_{0}+sn^{-1/3}}%
(t_{j}-t_{0})^{2}dH_{n}\\
&  \leq\text{var}(\psi^{\prime}(u))n^{1/3}s^{2}n^{-2/3}sn^{-1/3}\\
&  =\text{var}(\psi^{\prime}(u))s^{3}n^{-2/3}\\
&  =o(1)
\end{align*}
Then by (\ref{Lan5}) we obtain
\begin{equation}
\Lambda_{n}(s)\rightarrow^{p}\text{\textrm{E}$_{G}$}(\psi^{\prime}(u_{j}%
))\mu^{\prime}(t_{0})H^{\prime}(t_{0})\frac{s^{2}}{2}\text{ for }s>0.
\label{bnscac3}%
\end{equation}
Similarly we can prove that
\begin{equation}
\Lambda_{n}(s)\rightarrow^{p}\text{\textrm{E}$_{G}$}(\psi^{\prime}(u_{j}%
))\mu^{\prime}(t_{0})H^{\prime}(t_{0})\frac{s^{2}}{2}\text{ for }s<0.
\label{bnsac4}%
\end{equation}
Therefore from (\ref{nosac}), (\ref{Bns}), (\ref{bnsac1}), (\ref{bnsac2}),
(\ref{bnscac3}) and (\ref{bnsac4}) we get that%
\[
(s-r)\frac{\mathrm{\mathrm{E}_{G}}(\psi^{\prime}(u))}{\mathrm{\mathrm{E}_{G}%
}(\psi^{2}(u))^{1/2}}H^{\prime}(t_{0})^{1/2}\ n^{1/3}\tilde{\mu}_{n}%
^{c}(r,s)\overset{\mathcal{D}}{\Rightarrow}(B(s)-B(r))+\frac{\text{\textrm{E}%
$_{G}$}(\psi^{\prime}(u_{j}))}{\mathrm{\mathrm{E}_{G}}(\psi^{2}(u))^{1/2}}%
\mu^{\prime}(t_{0})H^{\prime}(t_{0})^{1/2}\frac{s^{2}-r^{2}}{2}\text{.}%
\]
Now the rest of the proof is as in Wright (1981).

\subsection{Proof of Theorem 2}

We require the following Lemma

\begin{lemma}
Assume A1-A5 Then,
\begin{equation}
\left\vert \frac{\partial}{\partial\sigma}\hat{\mu}_{n}(u,v,\sigma)\right\vert
\leq k,\text{ for all }u\leq v.
\end{equation}
\ \bigskip Proof
\end{lemma}

Taking the first derivative of Equation (\ref{eqn:zsigma}) with respect to
$\sigma$ yields
\[
\sum_{j\in C(u,v)}\psi^{\prime}\left(  \dfrac{x_{j}-\hat{\mu}_{n}(u,v,\sigma
)}{\sigma}\right)  \left\{  -\frac{1}{\sigma^{2}}(x_{j}-\hat{\mu}%
(u,v,\sigma))-\frac{1}{\sigma}\frac{\partial\hat{\mu}_{n}(u,v,\sigma
)}{\partial\sigma}\right\}  =0,
\]
and then%
\[
\frac{\partial}{\partial\sigma}\hat{\mu}_{n}(u,v,\sigma)=-\dfrac{\sum_{j\in
C(u,v)}\ \psi^{\prime}\left(  \dfrac{x_{j}-\hat{\mu}_{n}(u,v,\sigma)}{\sigma
}\right)  \dfrac{x_{j}-\hat{\mu}_{n}(u,v,\sigma)}{\sigma}}{\sum_{j\in
C(u,v)}\psi^{\prime}\left(  \dfrac{x_{j}-\hat{\mu}_{n}(u,v,\sigma)}{\sigma
}\right)  }.
\]
Let $D(u,v)=C(u,v)\cap\{j:|x_{j}-\hat{\mu}_{n}(u,v,\sigma)|/\sigma\leq k\}.$
Then by A5 we obtain%
\begin{align*}
\left\vert \frac{\partial}{\partial\sigma}\hat{\mu}_{n}(u,v,\sigma
)\right\vert  &  \leq\dfrac{\sum_{D}\psi^{\prime}\left(  \dfrac{x_{j}-\hat
{\mu}_{n}(u,v,\sigma)}{\sigma}\right)  \left\vert \dfrac{x_{j}-\hat{\mu}%
_{n}(u,v,\sigma)}{\sigma}\right\vert }{\sum_{D}\psi^{\prime}\left(
\dfrac{x_{j}-\hat{\mu}_{n}(u,v,\sigma)}{\sigma}\right)  }\\
&  \leq k.
\end{align*}
Therefore
\[
\left\vert \frac{\partial}{\partial\sigma}\hat{\mu}_{n}(u,v,\sigma)\right\vert
\leq k.
\]
\ 

\textbf{Proof of Theorem 2}

By the mean value theorem\
\[
\hat{\mu}_{n}(u,v,\hat{\sigma}_{n})=\hat{\mu}_{n}(u,v,\sigma_{0}%
)+\frac{\partial}{\partial\sigma}\hat{\mu}_{n}(u,v,\sigma_{n}^{\ast}%
)(\hat{\sigma}_{n}-\sigma),
\]
where $\sigma_{n}^{\ast}$ is some intermediate point between $\sigma$ and
$\hat{\sigma}_{n}$. Hence, by Lemma 2 we have
\begin{align*}
\max_{u\leq t}\min_{v\geq t}\hat{\mu}_{n}(u,v,\hat{\sigma}_{n})-k|\hat{\sigma
}_{n}-\sigma_{0}|  &  \leq\max_{u\leq t}\min_{v\geq t}\hat{\mu}_{n}%
(u,v,\sigma)\\
&  \leq\max_{u\leq t}\min_{v\geq t}\hat{\mu}_{n}(u,v,\hat{\sigma}_{n}%
)+k|\hat{\sigma}_{n}-\sigma_{0}|
\end{align*}
and A6 implies
\[
n^{1/3}|\hat{\mu}_{n}(t,\widehat{\sigma}_{n})-\hat{\mu}_{n}(t,\sigma_{0})|\leq
kn^{1/3}|\hat{\sigma}_{n}-\sigma_{0}|=o_{P}(1).
\]

\subsection{Proof of Theorem 3}

Without loss of generality we can assume that $\sigma_{0}=1.$ We consider
first the case $t^{\ast}=t_{0}.$ Assume that $x^{\ast}<\mu(t_{0})$. Then
$\delta_{(t_{0},x^{\ast})}$ represents a contamination model where an outlier
is placed at the observation point $t_{0}$ with value $x^{\ast}$ which is
below the trend $\mu(t_{0})$ at the point. \ Let $k=k(\varepsilon)$ be the
value such that
\[
T_{t_{0}}(\Lambda_{\varepsilon,t_{0},x^{\ast}})=m^{-}(t_{0},t_{0}%
-k,\Lambda_{0})=m(t_{0},t_{0}-k,t_{0},\Lambda_{0}).
\]
It is immediate that
\begin{equation}
\mu(t_{0}-k)=T_{t_{0}}(\Lambda_{\varepsilon})=m(t_{0},t_{0}-k,t_{0}%
,\Lambda_{0}). \label{1}%
\end{equation}
Then $m(t_{0}-k,t_{0},\Lambda_{0})$ should be the value of $m$ satisfying
\begin{equation}
\varepsilon\psi(x_{0}-m)+(1-\varepsilon)\int_{t_{0}-k(\varepsilon)}^{t_{0}%
}\int_{-\infty}^{\infty}\psi(\mu(t)+u-m)h(t)g(u)~dt~du=0, \label{cont}%
\end{equation}
and, since by (\ref{1}) $m=\mu(t_{0}-k),$ we have%
\begin{equation}
\varepsilon\psi(x_{0}-\mu(t_{0}-k(\varepsilon)))+(1-\varepsilon)\int
_{t_{0}-k(\varepsilon)}^{t_{0}}\int_{-\infty}^{\infty}\psi(\mu(t)+u-\mu
(t_{0}-k(\varepsilon))h(t)g(u)dtdu=0. \label{befapp}%
\end{equation}
Applying the Mean Value Theorem to the first term of (\ref{befapp}) we can
find $0\leq\varepsilon^{\ast}<\varepsilon$ such that%
\begin{align}
&  \psi(x^{\ast}-\mu(t_{0}-k(\varepsilon)))\nonumber\\
&  =\psi(x^{\ast}-\mu(t_{0}))-\psi^{\prime}(x^{\ast}-\mu(t_{0})-k(\varepsilon
^{\ast})\mu^{\prime}(t_{0}-k(\varepsilon^{\ast}))k(\varepsilon). \label{3}%
\end{align}

As for the second term in (\ref{befapp}) we also have that
\begin{align*}
&  \int_{t_{0}-k(\varepsilon)}^{t_{0}}\int\limits_{-\infty}^{\infty}\psi
(\mu(t)+u-\mu(t_{0}-k(\varepsilon))h(t)g(u)dtdu\\
&  =\int_{t_{0}-k(\varepsilon)}^{x^{\ast}}\int\limits_{-\infty}^{\infty}%
\psi(u)h(t)g(u)dtdu\\
&  +\int_{t_{0}-k(\varepsilon)}^{t_{0}}\int\limits_{-\infty}^{\infty}%
(\mu(t)-\mu(t_{0}-k(\varepsilon))g(u)h(t)\psi^{\prime}(u+\gamma)dudt,
\end{align*}
where $0\leq\gamma\leq\mu(t_{0})-\mu(t_{0}-k(\varepsilon))$. Since $\psi$ is
odd and $g$ even, $\int_{-\infty}^{\infty}\psi(u)g(u)du=0$, so that the first
term above vanishes. As for the second term, notice that
%
%\end{document}
%

\begin{align}
&  \int_{t_{0}-k(\varepsilon)}^{t_{0}}\int\limits_{-\infty}^{\infty}%
(\mu(t)-\mu(t_{0}-k(\varepsilon))g(u)h(t)\psi^{\prime}(u)dudt\nonumber\\
&  =\left[  \int_{t_{0}-k(\varepsilon)}^{t_{0}}(\mu(t)-\mu(t_{0}%
-k(\varepsilon))h(t)dt\right]  \left[  \int\limits_{-\infty}^{\infty}%
\psi^{\prime}(u+\gamma)g(u)du\right]  . \label{6}%
\end{align}
The first integral factor in the right hand side of the above display can be
further approximated. By the Mean Value Theorem, there exists $\xi(t)$ such
that $t_{0}-k(\varepsilon)\leq\xi(t)\leq t_{0}$ and
\begin{align}
\int_{t_{0}-k(\varepsilon)}^{t_{0}}(\mu(t)-\mu(t_{0}-k(\varepsilon))h(t)dt  &
=\int_{t_{0}-k(\varepsilon)}^{t_{0}}\mu^{\prime}(\xi(t))(t-t_{0}%
+k(\varepsilon))h(t)dt\nonumber\\
&  \approx\int_{t_{0}-k(\varepsilon)}^{t_{0}}\mu^{\prime}(t_{0})(t-t_{0}%
+k(\varepsilon))h(t)dt\nonumber\\
&  =\frac{\mu^{\prime}(t_{0})h(t_{0})}{2}\left[  (t-t_{0}+k(\varepsilon
))^{2}\right]  _{t_{0}-k(\varepsilon)}^{t_{0}}\nonumber\\
&  =\frac{1}{2}\mu^{\prime}(t_{0})h(t_{0})k^{2}(\varepsilon). \label{7}%
\end{align}

From expressions (\ref{3})-(\ref{7}) we obtain that Equation (\ref{befapp}),
can be written as
\[
\varepsilon\left[  \psi(x^{\ast}-\mu(t_{0}))-\psi^{\prime}(x^{\ast}-\mu
(t_{0})-k(\varepsilon^{\ast})\mu^{\prime}(t_{0}-k(\varepsilon^{\ast
}))k(\varepsilon)\right]  +(1-\varepsilon)\frac{1}{2}\mu^{\prime}%
(t_{0})h(t_{0})k^{2}(\varepsilon)\int\limits_{-\infty}^{\infty}\psi^{\prime
}(u)g(u+\gamma)du=0.
\]
Dividing both sides of this equation by $\varepsilon$ and using that
$k(\varepsilon)\rightarrow0$ $\ $\ and $\gamma\rightarrow0$ when
$\varepsilon\rightarrow0$ we obtain
\begin{equation}
\lim_{\varepsilon\rightarrow0}\frac{k^{2}(\varepsilon)}{\varepsilon}%
=-\frac{2\psi(x^{\ast}-\mu(t_{0}))}{h(t_{0})\mu^{\prime}(t_{0})\int
\limits_{-\infty}^{\infty}\psi^{\prime}(u)g(u)du}. \label{8}%
\end{equation}
Finally,\ according to (\ref{1}) \ and using the Mean Value Theorem, we can
write
\begin{align*}
\lim_{\varepsilon\rightarrow0}\frac{\left(  T_{t_{0}}(\Lambda_{\varepsilon
,t_{0},x^{\ast}})-T_{t_{0}}(\Lambda_{0})\right)  ^{2}}{\varepsilon}  &
=\lim_{\varepsilon\rightarrow0}\frac{(\mu(t_{0}-k(\varepsilon))-\mu
(t_{0}))^{2}}{\varepsilon}\\
&  =\lim_{\varepsilon\rightarrow0}\frac{\mu^{\prime2}(t^{\ast}(\varepsilon
))k^{2}(\varepsilon)}{\varepsilon},
\end{align*}
where $t^{\ast}(\varepsilon)\rightarrow t_{0}$. Then using equation (\ref{8})
we obtain that%
\begin{align*}
\text{IF}^{\ast}(T_{t_{0}},t_{0},x^{\ast})  &  =\lim_{\varepsilon\rightarrow
0}\frac{\left(  T_{t_{0}}(\Lambda_{\varepsilon,t_{0},x^{\ast}})-T_{t_{0}%
}(\Lambda_{0})\right)  ^{2}}{\varepsilon}\\
&  =-\dfrac{2\mu^{\prime}(t_{0})\psi(x_{0}-\mu(t_{0}))}{h(t_{0}%
)\text{\textrm{E}$_{G}$}(\psi^{\prime}(u))}\\
&  =\dfrac{2\mu^{\prime}(t_{0})\left\vert \psi(x_{0}-\mu(t_{0}))\right\vert
}{h(t_{0})\text{\textrm{E}$_{G}$}(\psi^{\prime}(u))}.
\end{align*}

The proof in the case the that $x^{\ast}<\mu(t_{0})$ is similar

We consider now the case $t^{\ast}>t_{0}.$ To prove this part of the theorem
is enough to show that\ there exists $\varepsilon^{\ast}>0$, so that
$\varepsilon\leq\varepsilon^{\ast}$ implies
\[
T_{t_{0}}(\Lambda_{\varepsilon,t^{\ast},x^{\ast}})=T_{t_{0}}(\Lambda_{0}%
)=\mu(t_{0}),
\]
and to prove this is enough to show that
\begin{equation}
\min_{s\geq0}m(t_{0},r,s,\Lambda_{\varepsilon,t^{\ast},x^{\ast}}%
)=m(t_{0},r,0,\Lambda_{\varepsilon,t^{\ast},x^{\ast}})=m(t_{0},r,0,\Lambda
_{0}). \label{impeq}%
\end{equation}
When $x^{\ast}\geq\mu(t_{0}),$ this is immediate. Consider the case that
$x^{\ast}<\mu(t_{0})$

Clearly for $0\leq s<t^{\ast}$%
\begin{align}
m(t_{0},r,s,\Lambda_{\varepsilon,t^{\ast},x^{\ast}})  &  =m(t_{0}%
,r,s,\Lambda_{0})\label{hinch1}\\
&  >m(t_{0},r,0,\Lambda_{0}).\nonumber
\end{align}
\ It is also easy to show that $s>t^{\ast}$ implies%
\begin{equation}
m(t_{0},r,s,\Lambda_{\varepsilon,t,x^{\ast}})\geq m(t_{0},r,t^{\ast}%
,\Lambda_{\varepsilon,t^{\ast},x^{\ast}}) \label{hinch2}%
\end{equation}
and for $r<0$ and \ for all $s$%
\begin{equation}
m(t_{0},r,s,\Lambda_{\varepsilon,t,x^{\ast}})<m(t_{0},0,s,\Lambda
_{\varepsilon,t,x^{\ast}}). \label{hinch3}%
\end{equation}
Then, using (\ref{hinch1})-(\ref{hinch3}) \ and the fact that $m(t_{0}%
,0,0,\Lambda_{0})=m(t_{0},0,0,,\Lambda_{\varepsilon,t^{\ast},x^{\ast}}),$ in
order to prove (\ref{impeq}), it is enough to show that
\begin{equation}
m(t_{0},0,t^{\ast},\Lambda_{\varepsilon,t^{\ast},x^{\ast}})>m(t_{0}%
,0,0,\Lambda_{0}). \label{enough1}%
\end{equation}
Recall that $m(t_{0},0,s,\Lambda_{\varepsilon,t^{\ast},x^{\ast}})$ is the
solution of
\[
\varepsilon\psi\left(  x^{\ast}-m\right)  1(t_{0}\leq t^{\ast}\leq
t_{0}+s)+(1-\varepsilon)V(s,m)=0,
\]
where%
\[
V(s,m)=\int_{-\infty}^{\infty}\int_{t_{0}}^{t_{0}+s}\psi\left(  \mu
(t)+u-m\right)  d\Lambda_{0}(t,u).
\]
Clearly $V(t^{\ast},m(t_{0},0,t^{\ast},\Lambda_{0}))=0$ and since
$m(r,t,\Lambda_{0})$ and $V(r,$ $t,m)$ are both increasing in $t$ we get
$V(r,t^{\ast},m(0,0,\Lambda_{0})<0.$Then, since $\psi$ is bounded, we can find
$\varepsilon^{\ast},$ so that for $\varepsilon<\varepsilon^{\ast}$ we have%
\[
\varepsilon\psi\left(  x^{\ast}-m\right)  1(t_{0}\leq t^{\ast}\leq
t_{0}+s)+(1-\varepsilon)V(s,m(0,0,\Lambda_{0}))<0,
\]
and therefore $m(0,t^{\ast},\Lambda_{\varepsilon,t^{\ast},x^{\ast}%
})>m(0,0,\Lambda_{0}).$ Then (\ref{enough1}) holds and this proves the Theorem
for the case $t^{\ast}>t_{0}.$ The proof for the case $t^{\ast}<t_{0}$ is similar.

\section{Proof of Theorem 4.}

Without loss of generality we can assume that $\sigma_{0}=1.$ It is easy to
see that the least favorable contaminating distribution is $\Lambda^{\ast}$
concentrated at $\delta_{t_{0},x_{0}}$ where $x_{0}$ tends to $-\infty$ or to
$\infty.$

A necessary and sufficient condition for $\varepsilon<\varepsilon^{\ast}$ is
that the equation
\begin{equation}
\varepsilon\psi(x_{0}-m)+(1-\varepsilon)\int_{0}^{t_{0}}\int_{-\infty}%
^{\infty}\psi(\mu(t)+u-m)h(t)g(u)dtdu=0 \label{BDP1}%
\end{equation}
have a bounded solution $m$\ solution for all $x_{0}<\mu(t_{0})$ and that the
equation
\begin{equation}
\varepsilon\psi(x_{0}-m)+(1-\varepsilon)\int_{t_{0}}^{1}\int_{-\infty}%
^{\infty}\psi(\mu(t)+u-m)h(t)g(u)dtdu=0 \label{BDP2}%
\end{equation}
have a solution for all $x_{0}>\mu(t_{0})$.

Taking $x_{0}\rightarrow-\infty$ we find that a sufficient condition for the
existence of a bounded solution of (\ref{BDP1}) for all $x_{0}<\mu(t_{0})$ is
that
\[
-\varepsilon k+(1-\varepsilon)kH(t_{0})\geq0,
\]
and this is equivalent to
\begin{equation}
\varepsilon\leq\frac{H(t_{0})}{1+H(t_{0})}. \label{epbd1}%
\end{equation}

Taking $x_{0}\rightarrow\infty$ we obtain that a sufficient condition for the
existence of solution of (\ref{BDP2}) for all $x_{0}>\mu(t_{0})$ is that
\[
\varepsilon k-(1-\varepsilon)k(1-H(t_{0}))\leq0,
\]
and this equivalent to%
\begin{equation}
\varepsilon\leq\frac{1-H(t_{0})}{2-H(t_{0})}. \label{epbd2}%
\end{equation}
The theorem follows from (\ref{BDP1}) and (\ref{BDP2}).\bigskip

\noindent{\Large \textbf{References}}\bigskip

%comment
%comment

\noindent Alvarez, E.E. y Dey, D.K. (2009), Bayesian Isotonic Changepoint
Analysis. \emph{Annals of the Institute of Statistical Mathematics}, 61, 355--370.

\noindent Brunk, H.D.\ (1958). On Estimation of Parameters Restricted by
Inequalities, \emph{Ann.\ Math.\ Statist.}, \textbf{29}, 437--454.

\noindent*Brunk, H.D.\ (1970) Estimation of Isotonic Regression,
\emph{Nonparametric Techniques in Statistical Inference}, 177--195, Cambridge Univ.\ Press.

\noindent De Boor, C.\ (2001) \emph{A Practical Guide to Splines}, Springer,
New York.

\noindent Dykstra, R.\ and Carolan, C.\ (1998) The Distribution of the
$\arg\max$ of two-sided Brownian Motion with Quadratic Drift. \emph{J.
Statist. Computat. Simulation,} \textbf{63}, 47--58.

\noindent Faraway, J.J.\ (2004). \emph{Linear Models with R}, Chapman \&
Hall/CRC, Boca Raton, FL.

\noindent Fomby, T.B. and Vogelsang, T.J.\ (2002) The application of
size-robust trend statistics to global-warming temperature series,
\emph{Journal of Climate}, \textbf{15}, 117--123.

\noindent Hampel, F. R. (1974). The influence curve and its role in robust
estimation. \emph{J. Amer. Statist. Assoc.},\textbf{69}, 383-393.

\noindent Ghement, I.R., Ruiz, M. and Zamar, R.H. (2008). Robust Estimation of
Error Scale in Nonparametric Regression Models. \emph{Journal of Statistical}
\emph{Planning and Inference}, \textbf{138}, 3200-3216.

\noindent Kulikova,V.N., and Lopuha\"{a}, H.P. (2006) The limit process of the
difference between the empirical distribution function and its concave
majorant, \emph{Statistics \& Probability Letters}, \textbf{76}, 1781--1786.

\noindent Maronna, R.A., Martin, R.D., and Yohai, V.J. (2006). \emph{Robust
Statistics Theory and Methods}, John Wiley, New York.

\noindent Meyer, M. (1996) Shape Restricted Inference with Applications to
Nonparametric Regression, Smooth Nonparametric Regression, and Density
Estimation, Ph.D. Thesis, Statistics, University of Michigan.

\noindent Pal, J.K. and Woodroofe, M. (2006) On the Distance Between
Cumulative Sum Diagram and Its Greatest Convex Minorant for Unequally Spaced
Design Points, \emph{Scand. J. Statist.} \textbf{33}, 279--291.

\noindent Prakasa Rao, B.L.S.\ (1969) Estimation of a Unimodal Density.
\emph{Sankhy\~{a} A}, \textbf{31}, 23--36.

\noindent Robertson, T., Waltman, P., (1968). On Estimating Monotone
Parameters. \emph{Ann. Math. Statist.}, \textbf{39}, 1030--1039.

\noindent Robertson, T., Wright, F.T. and Dyskra, R. L. (1988) \emph{Order
Restricted Statistical Inference}. New York John Wiley.

\noindent Shorack, G.\ R.\ (2000) \emph{Probability for Statisticians}.
Springer, New York.

\noindent Sun, J. and Woodroofe, M. (1999) Testing uniformity versus a
monotone density. \textit{Ann. Statist.} 27, 1, 338-360.

\noindent Wang, Y., and Huang, J., (2002) Limiting Distribution for Monotone
Median Regression. \emph{J.\ Statist.\ Plann.\ Inference}, \textbf{108}, 281--287.

\noindent Wright, F.T. (1981) The Asymptotic Behavior of Monotone Regression
Estimates. \emph{Ann. Statist.} \textbf{9}, 443--448.

\noindent Wu, W.B.; Woodroofe, M. and Mentz, G.B.\ (2001) Isotonic regression:
another look at the change-point problem. \emph{Biometrika} \textbf{88}, 793-804.

\end{document}